\title{Iterative Approximate Solutions of Kinetic Equations for Reversible Enzyme Reactions}
\author{S. Khoshnaw \\ Department of Mathematics, University of Leicester, LE1  7RH, UK \\  sarbazmath@yahoo.com or  sk464@le.ac.uk}
\documentclass[a4paper]{article}
\usepackage{chemarr, graphicx}  
\begin{document}    
 
\maketitle        
  
\begin{abstract} 
 We study kinetic models of reversible enzyme reactions and compare two techniques for analytic approximate solutions of the model. Analytic approximate solutions of non-linear reaction equations for reversible enzyme reactions are calculated using the Homotopy Perturbation Method (HPM) and the Simple Iteration Method (SIM). The results of the approximations are similar. The Matlab programs are included in appendices. 
\end{abstract} 


\section{Introduction} 
\label{sec:introduction}
The variety of chemical reactions in a living organism  are carried out by enzymes. It appears that the rate of chemical reactions (both forward and backward) are accelerated by enzymes. They are essential because many chemical reactions occur without the activity of enzymes. Such reactions are linked with an enzyme's active site, and they become a product after a series of stages. These stages are known as the enzymatic mechanism. There are two types of mechanisms, single substrate and multiple substrate mechanisms \cite{ Schnell2000, Maheswari2011, Varadharajan2011D, Varadharajan2011}. An important branch of enzymology is enzyme kinetics which is used to study the rate of chemical reactions. Differential equations are used to characterize the enzyme kinetics based on some principles of chemical kinetics \cite{Li2008, Rubinow1975, Murray1989, Segel1980}.

The single enzyme reaction is one of the most powerful kinds of kinetic reaction. Simply put, this enzyme reaction  is defined as follows:
\begin{equation}\label{equ:eq1}    
E+S \underset{k_{2}}{ \overset{k_{1}}{\rightleftharpoons}}  ES { \overset{k_{3}}{\longrightarrow}} P+E
\end{equation} 
\noindent where the concentrations of enzyme, substrate, enzyme-substrate complex and product are defined by $[E]$, $[S]$, $[ES]$ and $[P]$, respectively. Also, $k_{1}, k_{2}$ and $k_{3}$ represent the reaction rate constants. By using the idea of mass action, we can describe the reaction equation (\ref{equ:eq1}) in terms of a system of non-linear ordinary differential equations \cite{Varadharajan2011}.

There are varieties of possible simplifications for system (\ref{equ:eq1}) to describe analytic approximate solutions of the system. One of the most common approaches to simplifying this system is the use of quasi-steady state approximation (QSSA). The quasi-steady state assumptions occur as fundamental assumptions for enzyme kinetics, and the history of this subject began 80 years ago. It plays a key role with regard to the analysis of the enzyme kinetic equations \cite{Li2008}. Another simplification is the Michaelis-Menten equation created in 1913 which pointed out that the enzyme reaction (\ref{equ:eq1}) should be $k_{2} \gg k_{3}$, therefore $\frac{[E][S]}{[ES]} = \frac{k_{2}}{k_{1}}$. It  means that there is an equilibrium between $[E]$, $[S]$ and $[ES]$ to produce $[P]$ and $[E]$. In 1925, Briggs and Haldane proposed that the Michaelis-Menten assumption is not always applied. They said that it should be replaced by the assumption that $[ES]$ is present, not necessarily at equilibrium, but in a steady state under condition $[S_{0}] \gg [E_{0}]$. This means that the concentrations of  $[ES]$ occur as a steady state. This is known as the steady state assumption (SSA) or is sometimes called the quasi-steady state approximation (QSSA), or pseduo-steady sate approximation
\cite{Hanson2008}. 

\noindent The first description of QSS was given by Briggs and Haldane in 1925 \cite{Briggs1925}. They described the simplest enzyme reaction in equation (\ref{equ:eq1}), and pointed out the total concentration of enzyme $[E]$, where $[E]_{tot} = [E]+[ES]$ is a tiny value in comparison with the concentration of substrate $[S]$. Also, they have shown the term of $\frac{d[ES]}{dt}$ is negligible compared to $\frac{d[S]}{dt}$ and $\frac{d[P]}{dt}$. As a result, they found the Michaelis-Menten equation, which is a differential equation used to describe the rate of enzymatic reactions. The classical Michaelis-Menten equation is defined as, $k_{1}[E][S] = (k_{2}+k_{3})[ES]$, or
\begin{equation}\label{equ:eq2}
[ES] = \frac{[E][S]}{k_{M}+[S]}, \frac{d}{dt} [P] = k_{2}[ES] = \frac{k_{3}[E][S]}{k_{M}+[S]}
\end{equation}
\noindent where $k_{M} = \frac{k_{2}+k_{3}}{k_{1}}$ is the Michaelis-Menten constant (for more details see \cite{Gorban2011} ).

\noindent The purpose of this work is to derive asymptotic approximate expressions for the substrate, product, enzyme and enzyme-substrate concentrations for equation (\ref{equ:eq6}) by using (HPM) and (SIM), and to point out the similarities and differences between the methods of (HPM) and (SIM) for all values of dimensionless reaction diffusion parameters $\varepsilon$,  $\lambda$, $\alpha$ and $k$. Another aim of this project is to find out the appropriate iteration in (SIM) compared to (HPM).
\section{Mathematical Formulation }
The Michaelis - Meten equation (\ref{equ:eq1}) was applied by Kuhn in 1924 \cite{Meena2010} to several cases of enzyme kinetics. The model of biochemical reaction was developed by Briggs and Haldane in 1925 \cite{Varadharajan2011}. The model of an enzyme action considers a reaction that includes a substrate $[S]$ which binds an enzyme $[E]$ reversibly to a substrate-enzyme $[ES]$. The substrate- enzyme leads reversibly to product $[P]$ and enzyme $[E]$. This mechanism is often written as follows:
\begin{equation}\label{equ:eq6}
E+S \underset{k_{2}}{ \overset{k_{1}}{\rightleftharpoons}}  ES \underset{k_{4}}{ \overset{k_{3}}{\rightleftharpoons}} P+E 
\end{equation}
\noindent The mechanism shows the binding of substrate $[S]$ and the release of product $[P]$ where the free enzyme is $[E]$ and the enzyme-substrate complex is $[ES]$. In addition, $k _{1},  k_{2},  k_{3}$ and $k_{4}$ denote the rates of reaction. It is clear from equation (\ref{equ:eq6}) that substrate binding and product are reversible. The concentration of the reactants in equation (\ref{equ:eq6}) is denoted by lower case letters
\begin{equation} \label{equ:eq7}
e = [E], s = [S], c = [ES],  p = [P]
\end{equation}
\noindent The time of evolution of equations (\ref{equ:eq6}) and (\ref{equ:eq7}) are found by the law of mass action to obtain the set of system of the following non-linear reaction equations:
\begin{equation}\label{equ:eq8}
\frac{ds}{dt} = -k_{1} es + k_{2} c
\end{equation}
\begin{equation} \label{equ:eq9}
\frac{de}{dt} = -k_{1} es +( k_{2}+k_{3}) c-k_{4} pe
\end{equation}
\begin{equation}\label{equ:eq10}
\frac{dc}{dt} = k_{1} es -( k_{2}+k_{3}) c+k_{4} pe
\end{equation}
\begin{equation}\label{equ:eq11}
\frac{dp}{dt} = k_{3} c -k_{4} pe
\end{equation}
\noindent when the initial conditions at $t = 0$ are given by
\begin{equation}\label{equ:eq12}
e(0) = e_{0},  s(0) = s_{0},  c(0) = 0,  p(0) = 0
\end{equation}
\noindent Adding equations (\ref{equ:eq9}) and (\ref{equ:eq10}), and using initial conditions (\ref{equ:eq12}), we obtain
\begin{equation}\label{equ:eq13}
e+c = e_{0}
\end{equation}
\noindent Also, adding equations (\ref{equ:eq8}), (\ref{equ:eq10}) and (\ref{equ:eq11}), and using initial conditions (\ref{equ:eq12}), we get
\begin{equation} \label{equ:eq14}  
s+c+p = s_{0}
\end{equation}
\noindent By using equation (\ref{equ:eq13}) and equation (\ref{equ:eq14}), the system of ordinary differential equations (\ref{equ:eq8})-(\ref{equ:eq11}) reduce to only two variables, $s$ and $c$, as follows:
\begin{equation}\label{equ:eq15}
\frac{ds}{dt} = -k_{1} e_{0} s+(k_{1} s+k_{2}) c 
\end{equation}
\begin{equation}\label{equ:eq16}
\frac{dc}{dt} = k_{1} e_{0} s+(k_{1} s+k_{2}+k_{3}) c+k_{4}(e_{0}-c)(s_{0}-s-c)
\end{equation}
\noindent with initial conditions $s(0) = s_{0}$, $c(0) = 0$. By introducing the following parameters
\begin{equation} \label{equ:eq17}
\begin{split}
&\tau = \frac{k_{1} e_{0} t}{\varepsilon},  u(\tau) = \frac{s(t)}{s_{0}}, v(\tau) = \frac{c(t)}{e_{0}}, w(\tau) = \frac{p(t)}{e_{0}}, E(\tau) = \frac{e(0)}{e_{0}}, \\  &\lambda = \frac{k_{3}}{k_{1} s_{0}}, k = \frac{k_{2}+k_{3}}{k_{1} s_{0}},  \varepsilon = \frac{e_{0}}{s_{0}}, \alpha = \frac{k_{4}}{k_{1}}, m = \lambda+\alpha \varepsilon+\alpha
\end{split}
\end{equation}
\noindent we use the dimensionless technique to reduce the number of parameters for the system of equations (\ref{equ:eq15}) and (\ref{equ:eq16}) and the initial conditions (\ref{equ:eq12}). This can be represented in dimensionless form as follows:
\begin{equation}\label{equ:eq18}
\frac{du}{d\tau} = -\varepsilon u+ \varepsilon (u+k-\lambda)v
\end{equation}
\begin{equation}\label{equ:eq19}
\frac{dv}{d\tau} = u-(u+k)v+\alpha(1-v)(1-u-\varepsilon v)
\end{equation}
\begin{equation}\label{equ:eq20}
\frac{dw}{d\tau} = \alpha u-\alpha (uv+\varepsilon v^{2} +1)
\end{equation}
\begin{equation}\label{equ:eq21}
u(0) = 1, v(0) = 0,  w(0) = 0
\end{equation}
\noindent In this paper, we estimate the analytic approximate solution for a system of non-linear ODE equations (\ref{equ:eq18})-(\ref{equ:eq21}), by using the methods of (HPM) and (SIM).
\section{Analytical Approximate Solution using the Homotopy Perturbation Method }
The basic idea of the Homotopy-Perturbation Method (HPM) is defined in this section. It is then applied to find the approximate solution of the problem in equations (\ref{equ:eq18})-(\ref{equ:eq21}). It is considered from the following function:
\begin{equation}\label{equ:eq22}
A(x)-f(r) = 0,  r \in \Omega
\end{equation}
\noindent with the boundary conditions
\begin{equation}\label{equ:eq23}
B(x,\frac{\partial x}{\partial n}) = 0, r \in \Gamma
\end{equation}
\noindent where $A$, $B$, $f(r)$ and $\Gamma$ are general differential operators, boundary operators, a known analytic function, and the boundary of the domain $\Omega$, respectively \cite{He1999}. The function $A$ consists of linear part $L$ and non-linear part $N$. So, the equation (\ref{equ:eq22}) can be written as:
\begin{equation}\label{equ:eq24}
L(x)+N(x)-f(r) = 0
\end{equation}
\noindent The Homotopy function is defined by $z(r,q):\Omega \times [0,1] \rightarrow R
$, which satisfies
\begin{equation}\label{equ:eq25}
H(z,q) = (1-q)[L(z)-L(x_{0})]+q(A(z)-f(r)) = 0, q \in [0,1],  r \in \Omega,
\end{equation}
\noindent or,
\begin{equation}\label{equ:eq26}
H(z,q) = L(z)-L(x_{0})+qL(x_{0})+q[N(x)-f(r)]
\end{equation}
\noindent where $q \in[0,1]$ is an embedding parameter. At the same time, $x_{0}$ is an initial approximation of equation (\ref{equ:eq22}), which satisfies equation (\ref{equ:eq23}). Basically, from equation (\ref{equ:eq25}) and equation (\ref{equ:eq26}) we can obtain:
\begin{equation}\label{equ:eq27}
H(z,0) = L(z)-L(x_{0})=0,
\end{equation}
\begin{equation}\label{equ:eq28}
H(z,1) = A(z)-f(r) = 0 
\end{equation}
\noindent Changing $z(r,q)$ from $x_{0}$ to $x(r)$ depends on the values of $q$ from zero to unity. It is called deformation in the field of topology. At the same time, $L(z)-L(x_{0})$ and $A(z)-f(r)$ are called Homotopy. We use $q$ as a small parameter initially, and we defined the solution of equation (\ref{equ:eq25}) and equation (\ref{equ:eq26}) as a power series in $q$:
\begin{equation}\label{equ:eq29}  
z = z_{0}+q z_{1}+ q^{2} z_{2}+...
\end{equation}
\noindent Let $q=1$ to get the approximate solution of equation (\ref{equ:eq22}) 
\begin{equation}\label{equ:eq30} 
x = \lim_{q \to \ 1} \ z = z_{0}+ z_{1}+ z_{2}+... 
\end{equation}
Thus, HPM includes a combination of the perturbation method and the Homotopy method.
 Equations (\ref{equ:eq18})-(\ref{equ:eq20}) can be solved analytically in a simple and closed form by using the Homotopy Perturbation Method (HPM) in (Ref Appendix A). So, the approximate solutions of the system of non-linear differential equations (\ref{equ:eq18}) and (\ref{equ:eq19}) become:
\begin{equation}\label{equ:eq31}
\begin{split}
u(\tau) = &2 e^{-\varepsilon \tau}+\left(\frac{ab}{c-\varepsilon}+\frac{\alpha \varepsilon}{c}\right) t e^{-\varepsilon \tau}+ \frac{abc-a \alpha \varepsilon+a \alpha c}{c (\varepsilon-c)^{2}} e^{-c \tau}\\ &+ \frac{\alpha \varepsilon ^{2} -cb \varepsilon -c\alpha \varepsilon}{c^{2}(\varepsilon-c)} e^{(-\varepsilon-c) \tau}
+ \frac{a \alpha}{c\varepsilon}  +\frac{b}{\varepsilon-c} e^{-2 \varepsilon \tau} \\ &+\left(-1+\frac{a\alpha \varepsilon - abc -ac \alpha}{c (\varepsilon-c)^{2}} - \frac{a \alpha}{c \varepsilon}+\frac{b}{c-\varepsilon}+\frac{-\alpha \varepsilon^{2} + cb \varepsilon +c \alpha \varepsilon }{c^{2} (\varepsilon -c)} \right) e^{- \varepsilon \tau}
\end{split} 
\end{equation}


\begin{equation}\label{equ:eq32}
\begin{split}
v(\tau) = & \frac{b}{c-\varepsilon} e^{-\varepsilon \tau} + \frac{bc-\alpha \varepsilon +c \alpha}{c(\varepsilon-c)} e^{-c \tau} \\& +  (\frac{b}{c-\varepsilon}+\frac{b \alpha}{c(c-\varepsilon)}) e^{-\varepsilon \tau} + \frac{b^{2}}{(c-\varepsilon)(c-2 \varepsilon)} e^{-2 \varepsilon \tau}+ \frac{\alpha}{c } \\& +  \frac{\alpha \varepsilon b - c b^{2} -cb \alpha}{c \varepsilon (\varepsilon -c)} e^{(-\varepsilon -c) \tau} + ( \frac{b}{\varepsilon -c} \\& + \frac{b^{2}}{(\varepsilon -c)(c-2 \varepsilon)}+ \frac{c b^{2}-b \alpha \varepsilon + cb \alpha}{c \varepsilon(\varepsilon -c)}+ \frac{\alpha b}{c(\varepsilon -c)}) e^{-c \tau} 
\end{split}
\end{equation}
\noindent The analytic expressions of the substrate $u(\tau)$ and enzyme substrate $v(\tau)$ concentrations can be represented in equations (\ref{equ:eq31}) and (\ref{equ:eq32}). The dimensionless concentration of enzyme $E$ can be obtained from equations (\ref{equ:eq13}) and (\ref{equ:eq17}) as follows:
\begin{equation}\label{equ:eq33}
E(\tau) = \frac{e(t)}{e_{0}} = 1-v(\tau)
\end{equation}
\noindent The dimensionless concentration of the product $w$ is obtained either by equation (\ref{equ:eq20}) as follows:
\begin{equation} \label{equ:eq34}
w(\tau) = \int\limits_0^\tau (\alpha (u(t)- u(t)v(t)- \varepsilon v^{2}(t) -1)+mv(t))dt
\end{equation}
\noindent or we can use equation (\ref{equ:eq17}) and equation (\ref{equ:eq14}) to find the concentration of the product $w$ as follows:
\begin{equation}\label{equ:eq35}
w(\tau) = \frac{1-u(\tau)-\varepsilon v(\tau)}{\varepsilon}
\end{equation}


\noindent The simple analytic approximate solution form of the concentrations of enzyme $E(\tau )$ and product $w(\tau )$ for all values of parameters $\varepsilon$,  $\lambda$,  $\alpha$ and $k$, are represented in equations (\ref{equ:eq33})-(\ref{equ:eq35}). We can easily realize that the behaviour of the solution is based on this method (HPM) from Figures \ref{fig:fig4a}-\ref{fig:fig4e}. 

\section{Simple Iteration Method}
\label{sec:Simple Interaction Method}

In this section, we use a simple technique to find the analytic approximate solution for the system of equations (\ref{equ:eq18}) and (\ref{equ:eq19}). We introduce this method by rewriting equations (\ref{equ:eq18}) and (\ref{equ:eq19}) as follows:
\begin{equation}\label{equ:eq36}
\frac{du}{d \tau} = - \varepsilon u+\varepsilon(k-\lambda)v+\varepsilon uv
\end{equation}
\begin{equation}\label{equ:eq37}
\frac{dv}{d \tau} = (1-\alpha) u-(k+\alpha+\alpha \varepsilon) v - (1-\alpha) uv+ \alpha \varepsilon v^{2} +\alpha
\end{equation}
\noindent Let $a = \varepsilon (k-\lambda)$,  $b = (1-\alpha)$ and $c = k+\alpha+\alpha \varepsilon$, then the equations (\ref{equ:eq36}) and (\ref{equ:eq37}) can be written as:
\begin{equation}\label{equ:eq38}
 \\\left( \begin{array}{ccc}
u^\prime _{n+1} \\
v^\prime _{n+1} \\
\end{array} \right)\ = A \left( \begin{array}{ccc}
u _{n+1} \\
v _{n+1} \\
\end{array} \right)\  +G(u_{n}, v_{n})
\end{equation}
\noindent for $n = 0, 1, 2, ...$ 
\noindent where, $ G(u_{n}v_{n}) = \left( \begin{array}{ccc}
\varepsilon u_{n} v_{n} \\
-b u_{n} v_{n}+\alpha \varepsilon v^{2}_{n}+\alpha \\
\end{array} \right)\ $ is a non-linear part of system (\ref{equ:eq38}), and

$A = \left( \begin{array}{ccc}
-\varepsilon &  a \\
b   &  -c \\
\end{array} \right)\ $ is a matrix of linear part of system (\ref{equ:eq38}). To evaluate an approximate solution of equation (\ref{equ:eq38}) with the initial conditions implied by equation (\ref{equ:eq21}), we introduce the following steps to approach the approximate solution:
\paragraph{Step 1.}For $n = 0,$ $u_{0}(\tau) = 1, v_{0}(\tau) = 0$  and, if possible suppose that $\alpha \longrightarrow 0$ (just in this step). It means we assume the non-linear part of equation (\ref{equ:eq38}) approaches zero. Consequently, we obtain the following system:
\begin{equation}\label{equ:eq39}
 \\\left( \begin{array}{ccc}
u^\prime _{1} \\
v^\prime _{1} \\
\end{array} \right)\ = A \left( \begin{array}{ccc}
u _{1} \\
v _{1} \\
\end{array} \right)\
\end{equation}
\noindent We can solve the system of ordinary differential equations (\ref{equ:eq39}) analytically \cite{Dawkins2007}. So, the solution of equation (\ref{equ:eq39}) with initial conditions (\ref{equ:eq21}) is
\begin{equation}\label{equ:eq40}
 \\\left( \begin{array}{ccc}
u _{1} (\tau) \\
v _{1} (\tau) \\
\end{array} \right)\ = \left( \begin{array}{ccc}
d_{2} e^{p_{1} \tau }+d_{3} e^{p_{2} \tau }  \\
d_{1} e^{p_{1} \tau }-d_{1} e^{p_{2} \tau }  \\
\end{array} \right)\
\end{equation}
\noindent where $p_{1} $ and $p_{2}$ are eigenvalues of matrix $A$, and $d_{1} = \frac{(p_{1}+\varepsilon)(p_{2}+\varepsilon)}{a (p_{2}-p_{1})} $, $d_{2} = \frac{(p_{2}+\varepsilon)}{a (p_{2}-p_{1})}$, and $ d_{3} = \frac{(p_{1}+\varepsilon)}{ (p_{1}-p_{2})}$. We substitute $u_{1}$ and $v_{1}$ in equations (\ref{equ:eq33}) and (\ref{equ:eq35}), then obtain $E_{1}$ and $w_{1}$, respectively. The behaviour of components  in equation (\ref{equ:eq40}) are described in Figures \ref{fig:fig1a}-\ref{fig:fig1e} (see Appendix C).

\paragraph{Step 2.}For $n = 1$, and substituting equation (\ref{equ:eq40}) in equation (\ref{equ:eq38}), we obtain the following system of non-linear ODE:
\begin{equation}\label{equ:eq41}
 \\\left( \begin{array}{ccc}
u^\prime _{2} \\
v^\prime _{2} \\
\end{array} \right)\  = A \left( \begin{array}{ccc}
u _{2} \\
v _{2} \\
\end{array} \right)\  +G(u_{1}, v_{1})
\end{equation}
\noindent It is clear that the system of non-linear differential equations (\ref{equ:eq41}) is solved analytically \cite{Dawkins2007}. The solution of the system with initial conditions (\ref{equ:eq21}) is obtained as follows: 
\begin{equation}\label{equ:eq42}
u _{2} (\tau) = ac_{3} e^{p_{1} \tau } +ac_{4} e^{p_{2} \tau } +d_{22} e^{ 2p_{1} \tau } +d_{23} e^{(p_{1}+p_{2}) \tau }+d_{24} e^{2p_{2} \tau }+d_{25}
\end{equation}
\begin{equation}\label{equ:eq43}
v _{2} (\tau) = c_{3}(p_{1}+\varepsilon) e^{p_{1} \tau } +c_{4}(p_{2}+\varepsilon) e^{p_{2} \tau } +d_{26} e^{ 2p_{1} \tau } +d_{27} e^{(p_{1}+p_{2}) \tau }+d_{28} e^{2p_{2} \tau }+d_{29}
\end{equation}
\noindent where $d_{22}, ..., d_{29}$ and $c_{3}, c_{4}$ are constants. We substitute $u_{2}$ and $v_{2}$ in equations (\ref{equ:eq33}) and (\ref{equ:eq35}), and obtain $E_{2}$ and $w_{2}$, respectively. The behaviour of concentrations in this step is described in Figures \ref{fig:fig2a}-\ref{fig:fig2e} (See Appendix D).

\paragraph{Step 3.}For $n = 2$, and substituting equations (\ref{equ:eq42}) and (\ref{equ:eq43}) in equation (\ref{equ:eq38}), we get the following system of non-linear ODE,
\begin{equation}\label{equ:eq44}
 \\\left( \begin{array}{ccc}
u^\prime _{3} \\ 
v^\prime _{3} \\
\end{array} \right)\  = A \left( \begin{array}{ccc}
u _{3} \\
v _{3} \\
\end{array} \right)\  +G(u_{2}, v_{2})
\end{equation}
\noindent The system of non-linear differential equations (\ref{equ:eq44}) is solved analytically. The solution of the system with initial conditions (\ref{equ:eq21}) is obtained as follows:
\begin{equation}\label{equ:eq45}
\begin{split}
u _{3} (\tau) = & ac_{5} e^{p_{1} \tau } +ac_{6} e^{p_{2} \tau } +d_{90} e^{ 2p_{1} \tau } +d_{91} e^{(p_{1}+p_{2}) \tau }+d_{92} e^{3p_{1} \tau } \\&+ d_{93} e^{(2p_{1}+p_{2}) \tau }+d_{94} e^{(p_{1}+2p_{2}) \tau }+d_{95} \tau e^{ p_{1} \tau }+d_{96} e^{ p_{1} \tau} \\&+ d_{97} e^{ 2p_{2} \tau }+d_{98} e^{ 3p_{2} \tau } +d_{99} e^{ p_{2} \tau }+d_{100} \tau  e^{ p_{2} \tau }+d_{101} e^{ 4p_{1} \tau} \\&+ d_{102} e^{(3p_{1}+p_{2}) \tau }+d_{103} e^{(2p_{1}+2p_{2}) \tau }+d_{104} e^{(p_{1}+3p_{2}) \tau }+d_{105} e^{ 4p_{2} \tau }+d_{106}
\end{split}
\end{equation}
\begin{equation}\label{equ:eq46}
\begin{split}
v _{3} (\tau) = & c_{5} h_{1} e^{p_{1} \tau } +c_{6} h_{2} e^{p_{2} \tau } +d_{107} e^{ 2p_{1} \tau } +d_{108} e^{(p_{1}+p_{2}) \tau} \\&+  d_{109} e^{3p_{1} \tau }+d_{110} e^{(2p_{1}+p_{2}) \tau }+d_{111} e^{(p_{1}+2p_{2}) \tau }+d_{112} \tau e^{ p_{1} \tau } \\&+ d_{113} e^{ p_{1} \tau }+d_{114} e^{ 2p_{2} \tau }+d_{115} e^{ 3p_{2} \tau } +d_{116} e^{ p_{2} \tau } \\&+ d_{117} \tau  e^{ p_{2} \tau }+  d_{118} e^{ 4p_{1} \tau } + d_{119} e^{(3p_{1}+p_{2}) \tau }+d_{120} e^{(2p_{1}+ 2p_{2}) \tau } \\&+ d_{121} e^{(p_{1}+3p_{2}) \tau }+d_{122} e^{ 4p_{2} \tau }+d_{123}
\end{split}
\end{equation} 
\noindent where $d_{90},  ...,  d_{123} $ and $c_{5}, c_{6}$ are constants. We substitute $u_{3}$ and $v_{3}$ in equations (\ref{equ:eq33}) and (\ref{equ:eq35}), and obtain $E_{3}$ and $w_{3}$, respectively. The behaviour of concentrations of this iteration is described in Figures \ref{fig:fig3a}-\ref{fig:fig3e} (see Appendix E).
\section{Asymptotic Analysis}
\label{sec:Asy Analy}

An important development of asymptotic analysis was suggested by Kruskal (1963) for differential equations  \cite{Gorban2010}. He defined asymptotology as '' the art of describing the behaviour of specified solution (or family of solutions) of a system in limiting case''. The following three different conditions can be identified based on the initial ratio $\frac{[E_{0}]}{[S_{0}]}$ \cite{Kargi2009}.

\noindent a) If the initial concentration of enzyme $[E_{0}]$ is much greater than the initial concentration of substrate $[S_{0}]$. This means that $\frac{[E_{0}]}{[S_{0}]} \gg 1$. Also, Schenell and Maini in \cite{Schnell2000} emphasised that the initial concentration of enzyme greatly exceeds the concentration of substrate, that is $[E_{0}] \gg [S_{0}]$.  So, from equation (\ref{equ:eq17}), we get $\varepsilon > 1$.  In this case, the part of the enzyme concentration which binds to the concentration of the substrate is small. This means that there is a free rate of enzyme. This rate is based on the availability of the substrate, and is increased whenever the concentrations of substrate are increased, or by adding additional substrate to the chemical reaction.

\noindent b) If the initial concentration of substrate $[S_{0}]$ is much greater than the initial concentration of enzyme $[E_{0}]$. This means that $\frac{[E_{0}]}{[S_{0}]} \ll 1$.  So, from equation (\ref{equ:eq17}), we obtain $\varepsilon < 1$. In this case, there is a small part of substrate that links to the enzyme, while a part of it is free. In this case, enzyme molecules usually bind to substrate molecules which means that a small amount of enzyme  is free. The availability of enzyme in this case depends on this rate, and increases when the rate of enzyme is increased, or by adding some extra enzyme to the chemical reaction.

\noindent c) If the initial concentration of enzyme $[E_{0}]$ and substrate are equal. This means $\frac{[E_{0}]}{[S_{0}]} =1$, so from equation (\ref{equ:eq17}), we get $\varepsilon = 1$. In this case, there are no any free molecules of enzyme or substrate. In other words, all substrate molecules are occupied by the enzyme molecules, and all enzyme molecules are also limited by the number molecules of the substrate.

\noindent Furthermore, if we look the constant rate of reactions $k_{4}$ and $k_{1}$ from equation (\ref{equ:eq17}), we can define the following conditions:

d. If $ k_{4} \gg k_{1} $ then $ \alpha >1 $ .

e. If $ k_{4} \ll k_{1} $ then $ \alpha <1 $ .

f. If $ k_{4} \sim k_{1} $ then $ \alpha \approx 1 $ .

\noindent In addition,  according to the definition of $\lambda$ and $k$ from equation (\ref{equ:eq17}), we obtain $\lambda < k$, because $k_{3}$ always has a positive value. As result, we can easily combine the conditions (a)-(f). We then get the following five basic cases considered in this paper:

Case 1. The value of $\varepsilon  \sim 1$ and $\alpha \sim 1$.

Case 2. The value of $\varepsilon > 1$ and $ \alpha > 1$.

Case 3. The value of $\varepsilon < 1$ and $\alpha < 1$.

Case 4. The value of $\varepsilon > 1$ and $\alpha < 1$.

Case 5. The value of $ \varepsilon < 1$ and $\alpha > 1$.

\noindent We apply the above cases separately in the analytic approximate solution for both methods (HPM) and (SIM). 
\section{Results and discussions }
\label{sec:Results and diss}

\begin{figure}[h]
\center 
\includegraphics[width=0.5\textwidth]{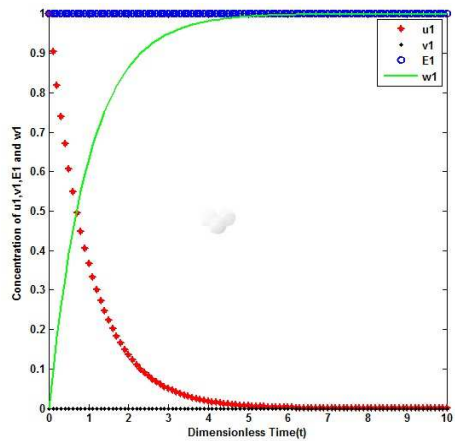}
\caption{ $\varepsilon = 1, \alpha = 1, \lambda = 0.4$ and $k = 1.3$}
\label{fig:fig1a}
\end{figure}

\begin{figure}[h]
\center
\includegraphics[width=0.6\textwidth]{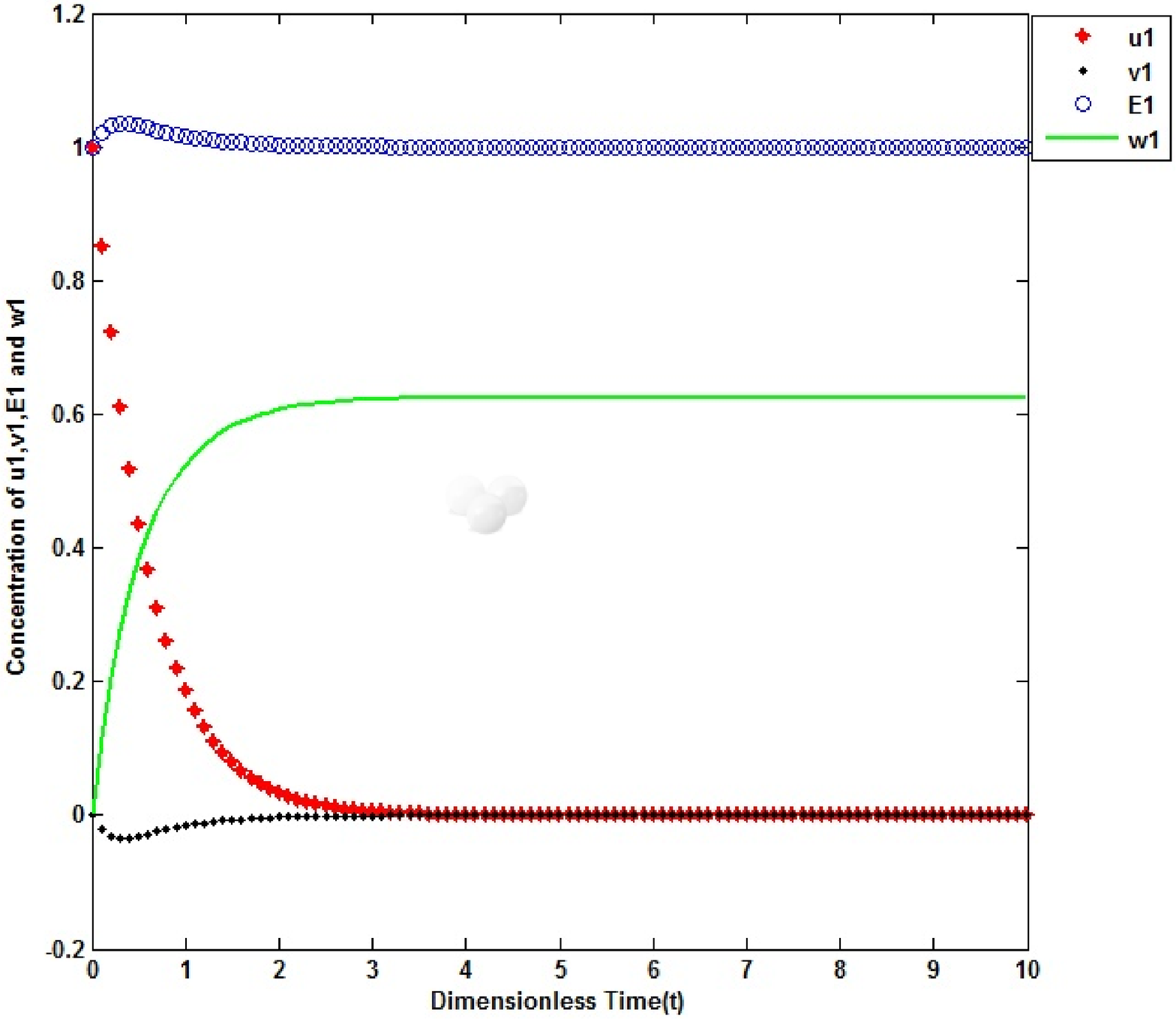}
\caption{ $\varepsilon = 1.6, \alpha = 1.3, \lambda = 0.9$ and $k = 1.7$}
\label{fig:fig1b}
\end{figure}

\begin{figure}[h]
\center
\includegraphics[width=0.55\textwidth]{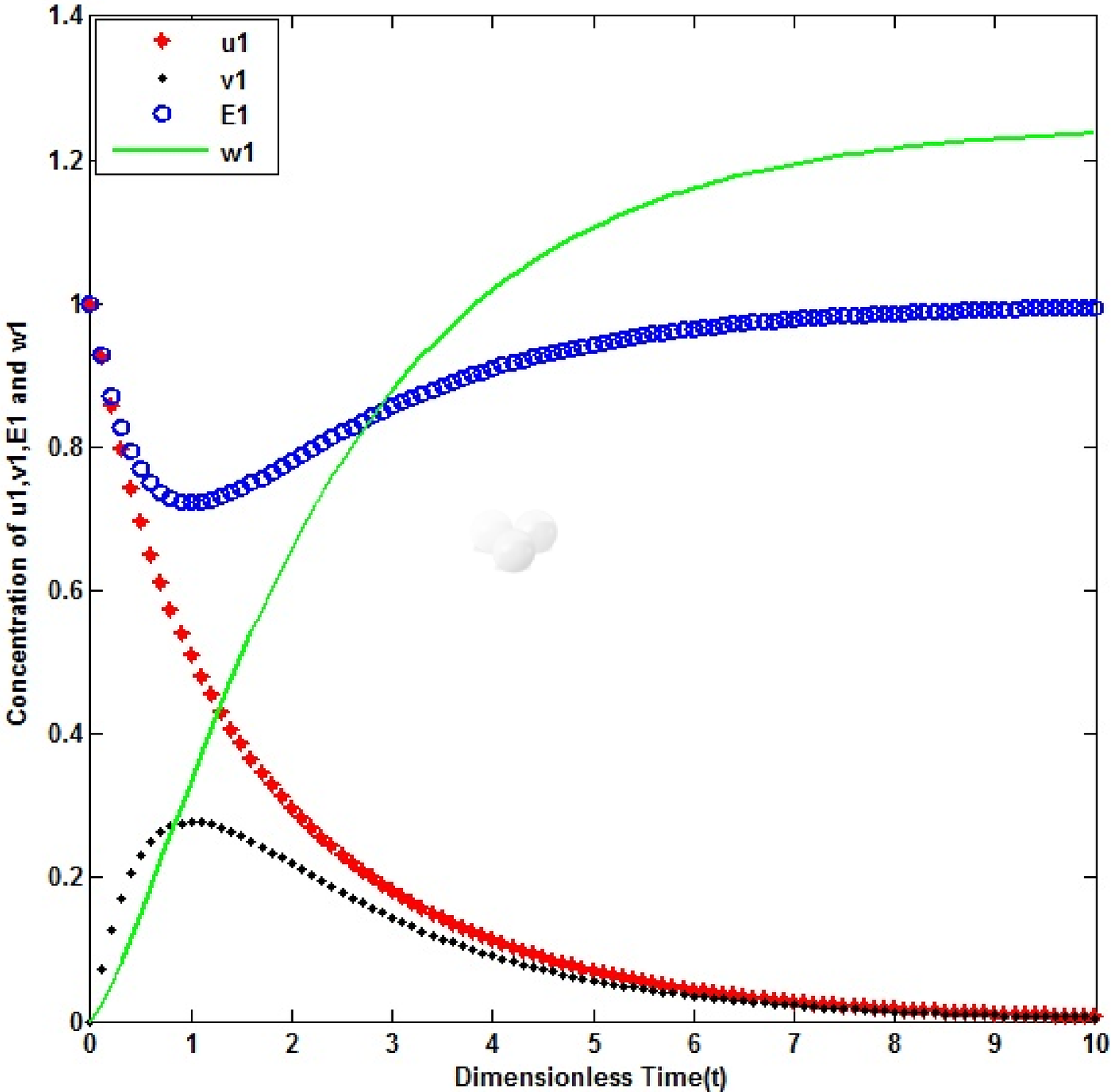}
\caption{$\varepsilon = 0.8, \alpha = 0.2, \lambda = 0.6$ and $k = 1.1$}
\label{fig:fig1c}
\end{figure}

\begin{figure}[h]
\center
\includegraphics[width=0.6\textwidth]{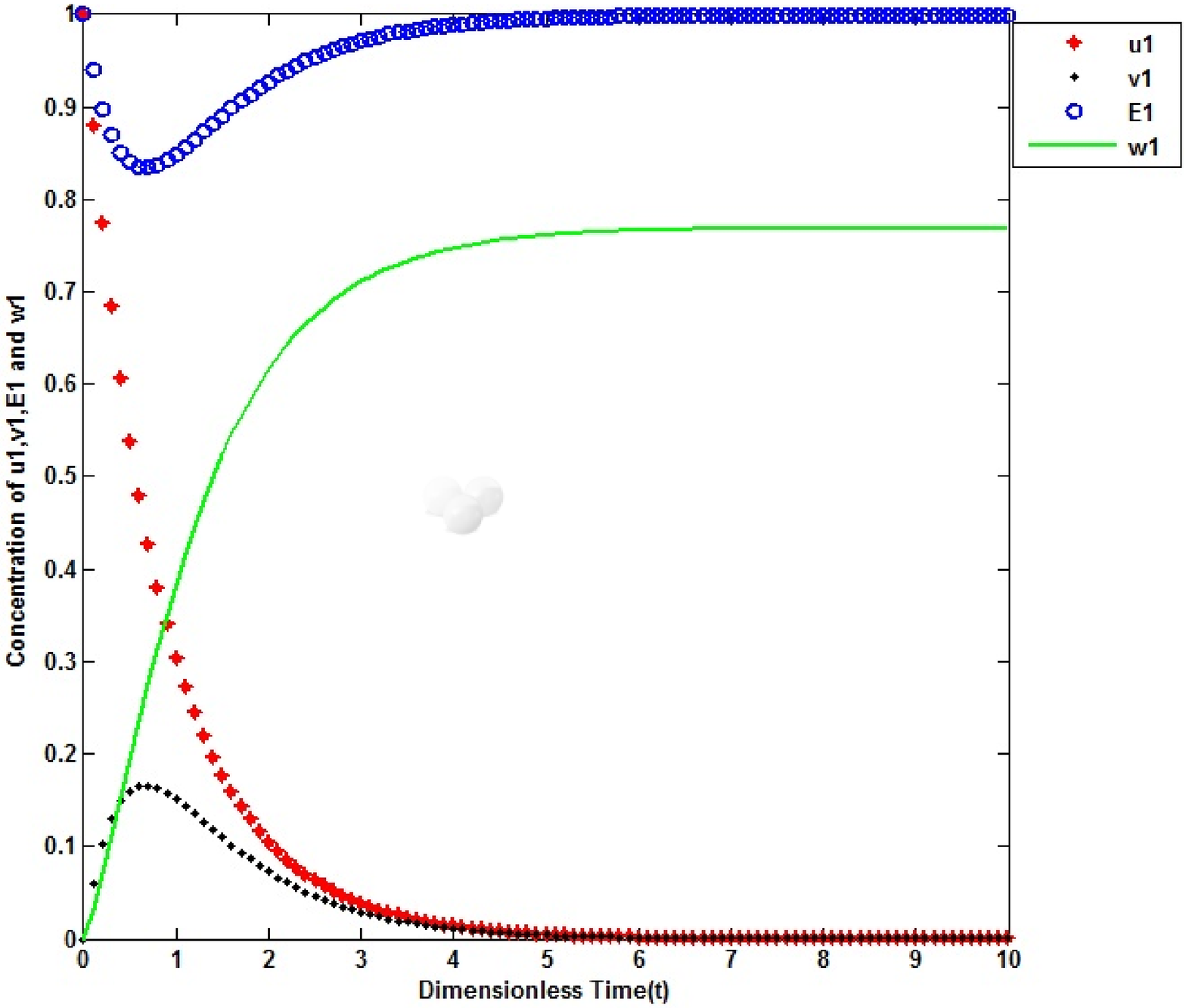}
\caption{$\varepsilon = 1.3, \alpha = 0.3, \lambda = 0.9$ and $k = 1.2$}
\label{fig:fig1d}
\end{figure}

\begin{figure}[h]
\center
\includegraphics[width=0.52\textwidth]{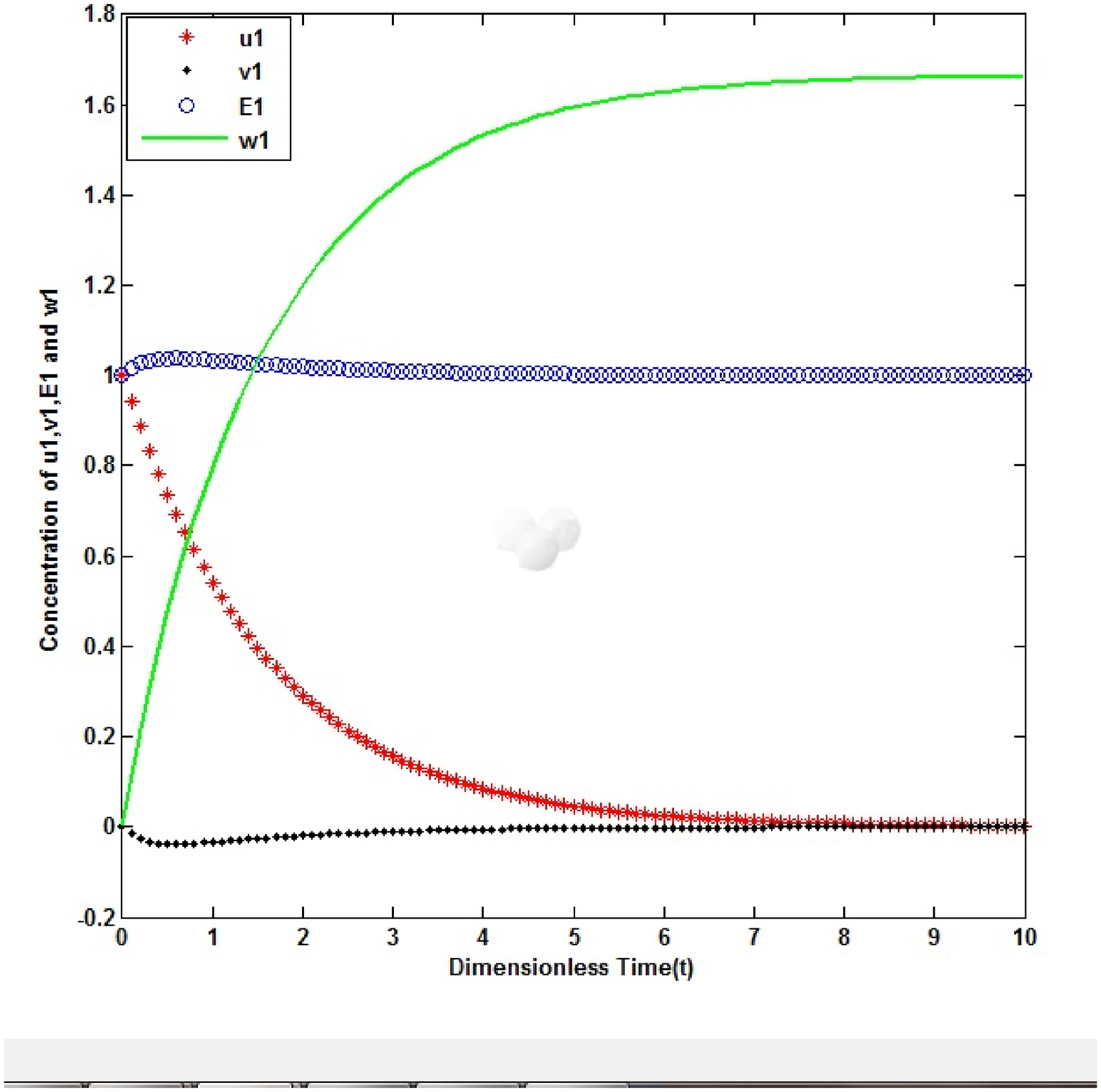}
\caption{$\varepsilon = 0.6, \alpha = 1.2, \lambda = 1.2$ and $k = 1.7$}
\label{fig:fig1e} 
\end{figure}

\begin{figure}[h]
\center 
\includegraphics[width=0.52\textwidth]{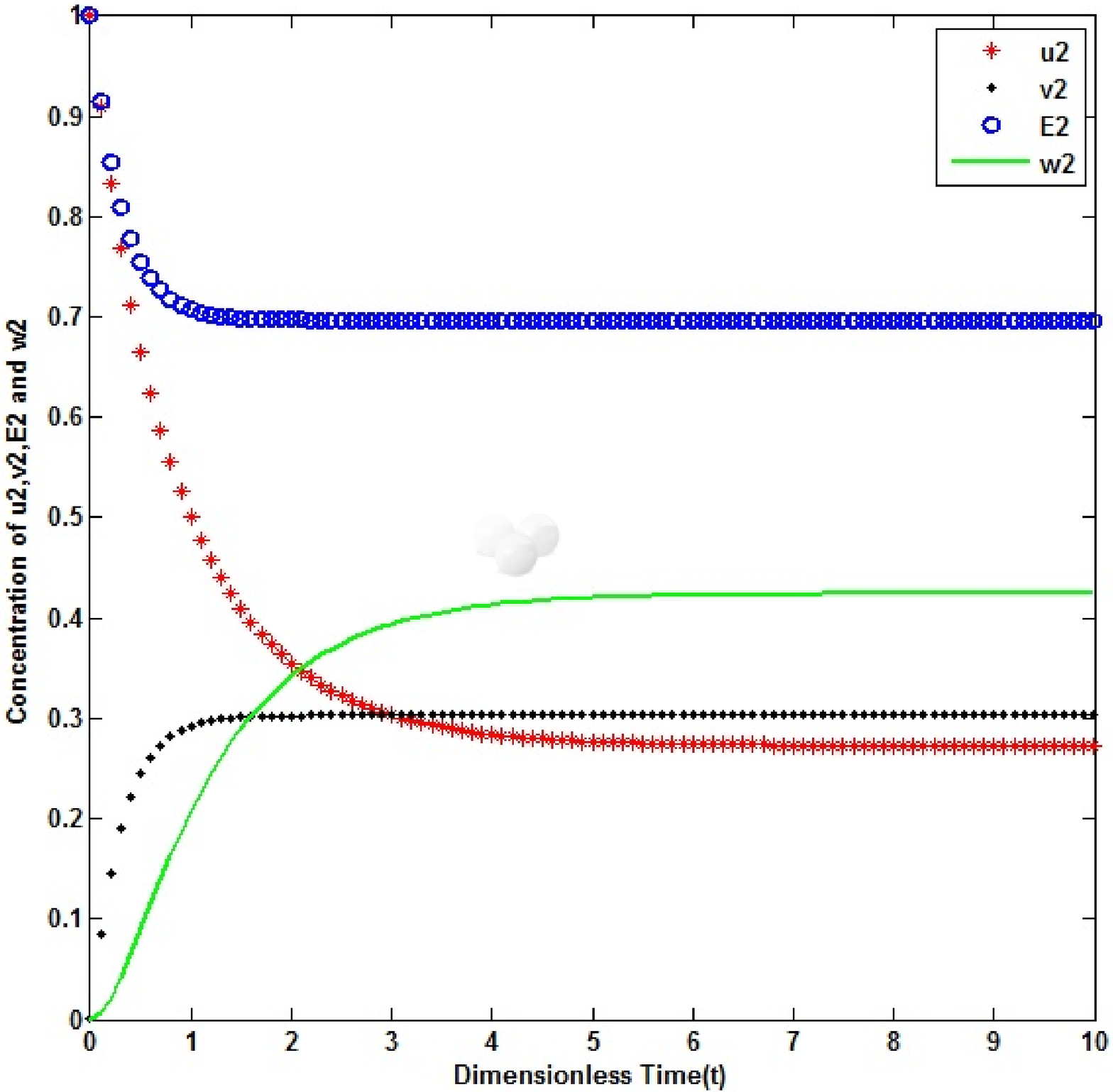}
\caption{ $\varepsilon = 1, \alpha = 1, \lambda = 0.4$ and $k = 1.3$}
\label{fig:fig2a}
\end{figure}

\begin{figure}[h]
\center
\includegraphics[width=0.57\textwidth]{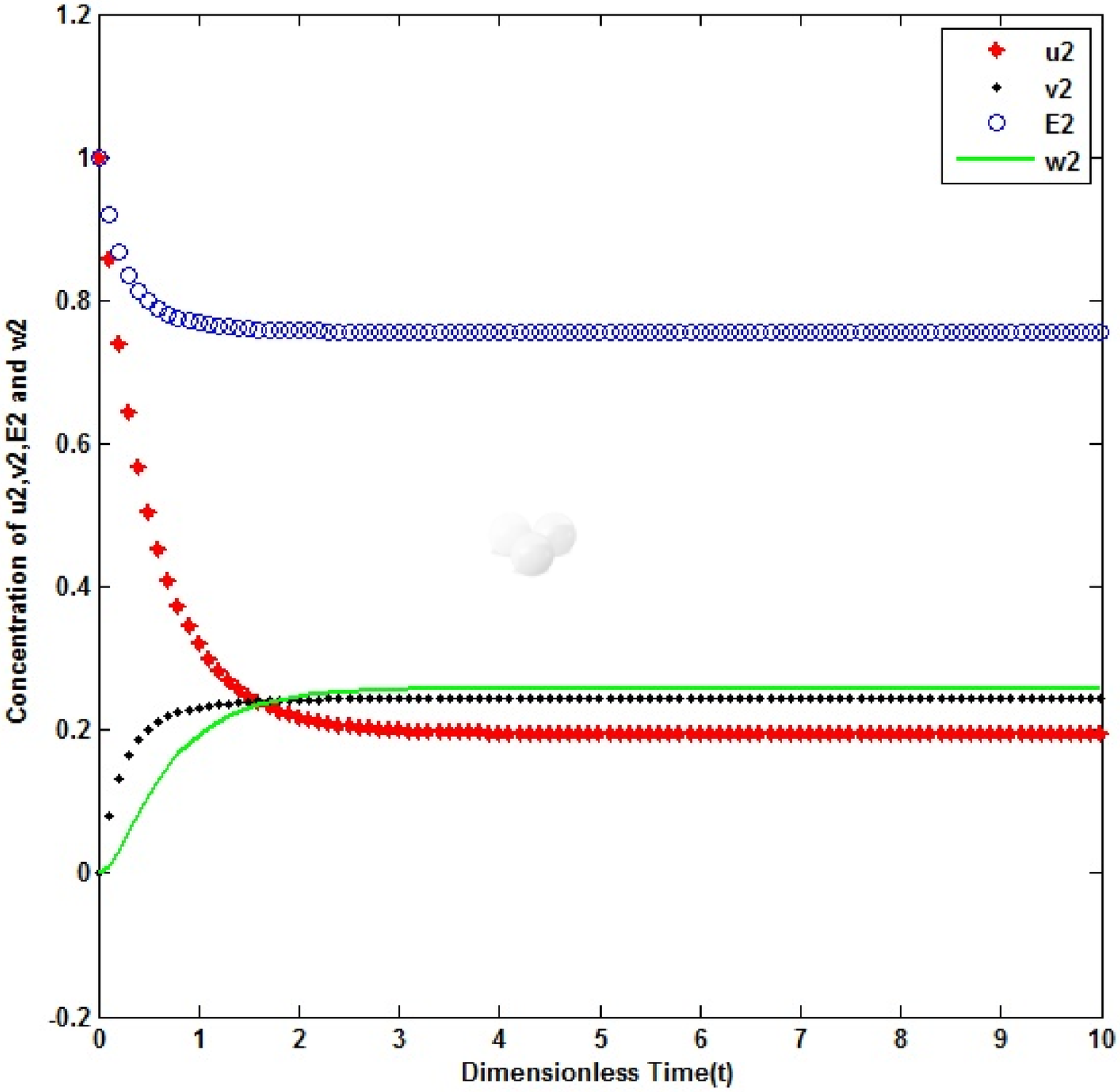}
\caption{ $\varepsilon = 1.6, \alpha = 1.3, \lambda = 0.9$ and $k = 1.7$}
\label{fig:fig2b}
\end{figure}

\begin{figure}[h]
\center
\includegraphics[width=0.57\textwidth]{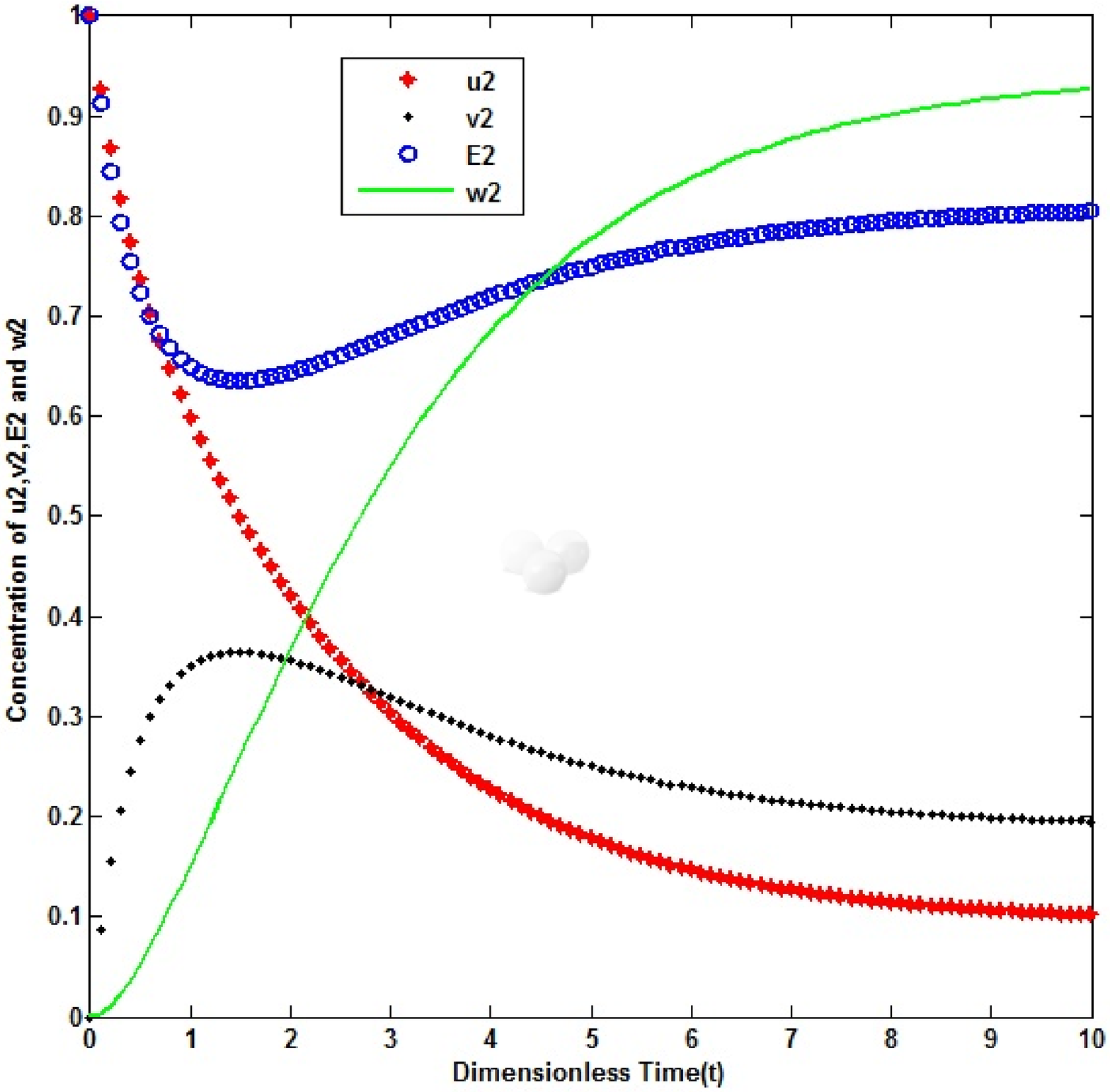}
\caption{$\varepsilon = 0.8, \alpha = 0.2, \lambda = 0.6$ and $k = 1.1$}
\label{fig:fig2c}
\end{figure}

\begin{figure}[h]
\center
\includegraphics[width=0.57\textwidth]{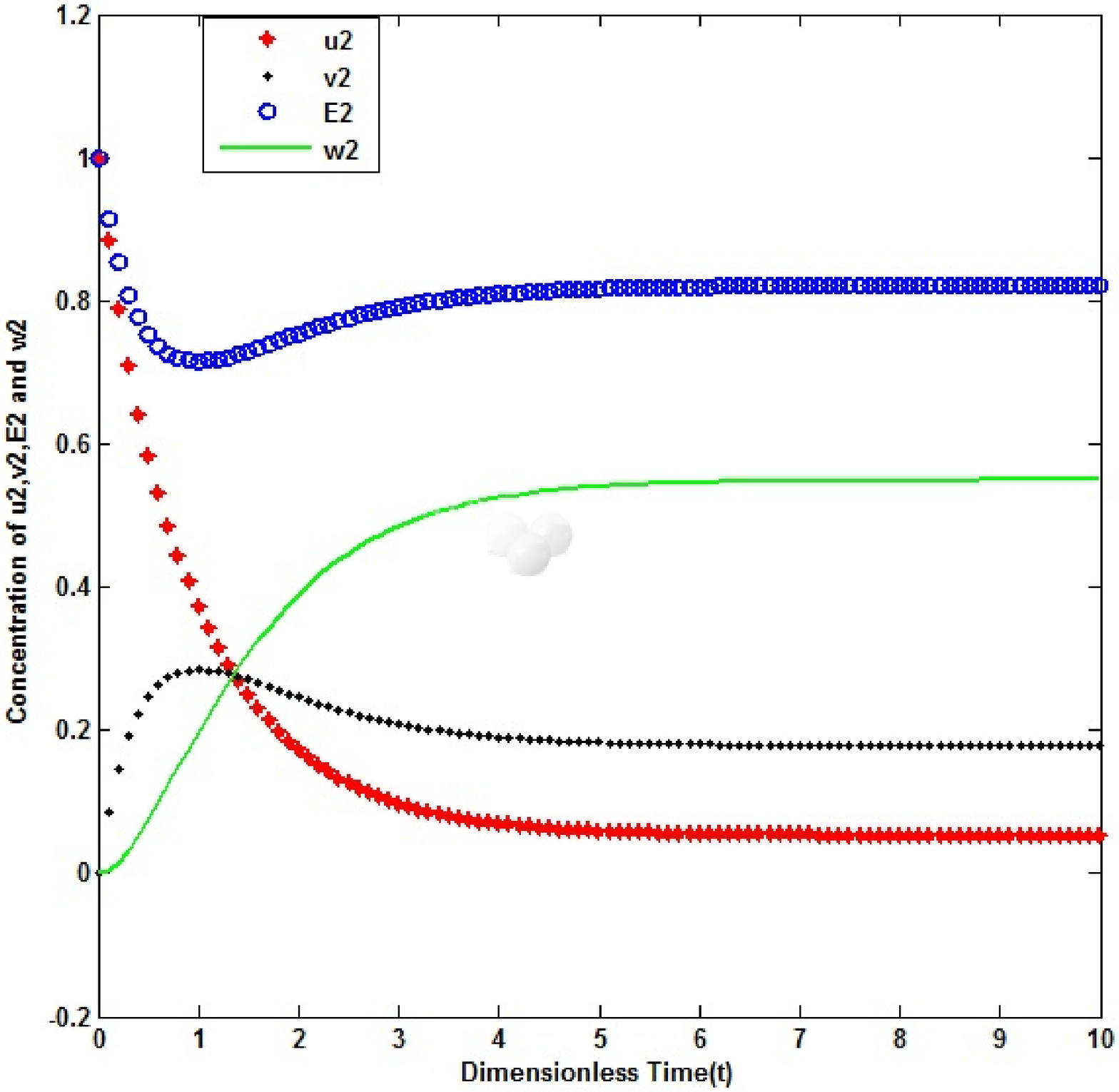}
\caption{ $\varepsilon = 1.3, \alpha = 0.3, \lambda = 0.9$ and $k = 1.2$}
\label{fig:fig2d}
\end{figure}

\begin{figure}[h]
\center
\includegraphics[width=0.57\textwidth]{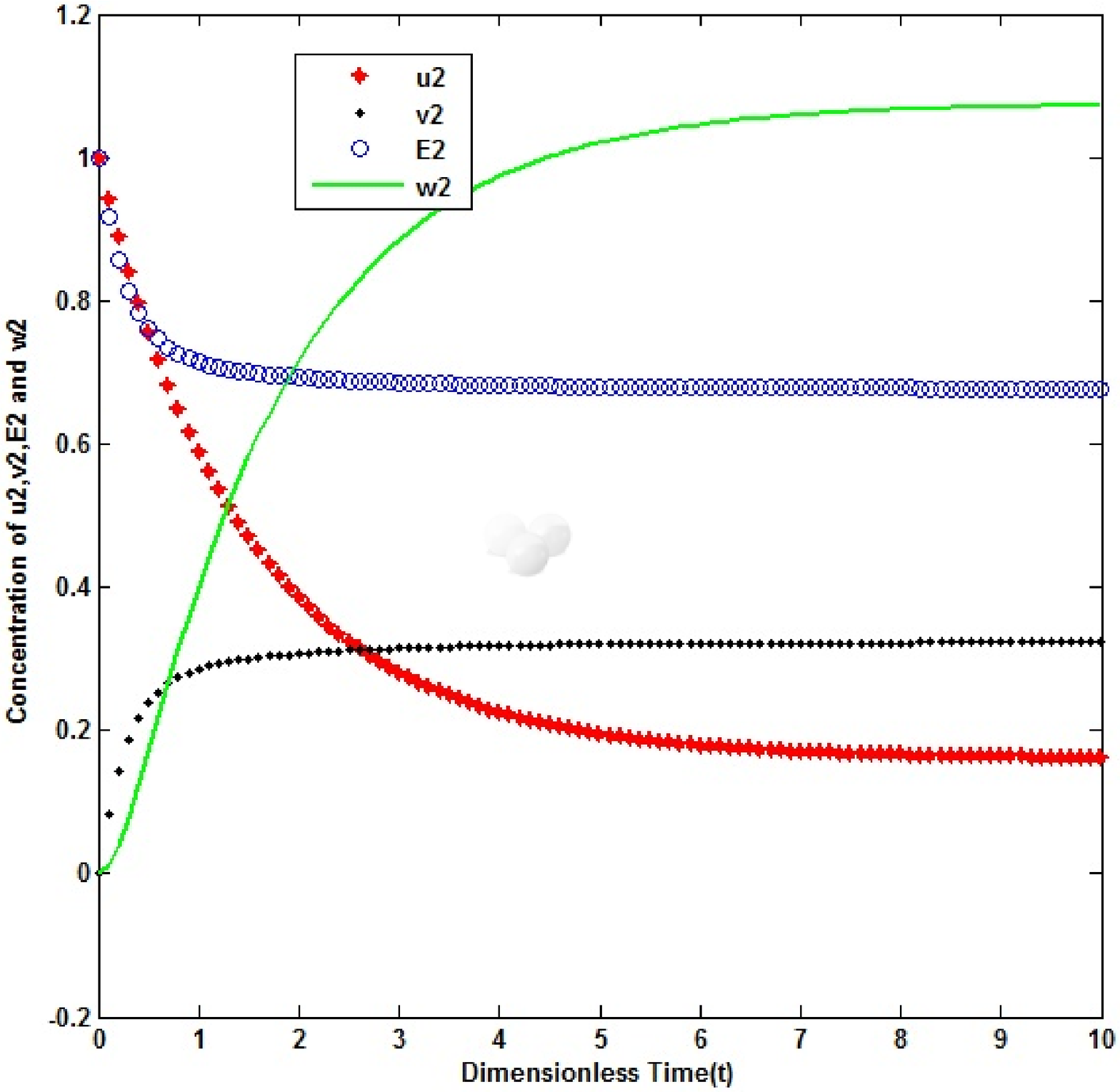}
\caption{ $\varepsilon = 0.6, \alpha = 1.2, \lambda = 1.2$ and $k = 1.7$}
\label{fig:fig2e} 
\end{figure}

\begin{figure}[h]
\center 
\includegraphics[width=0.5\textwidth]{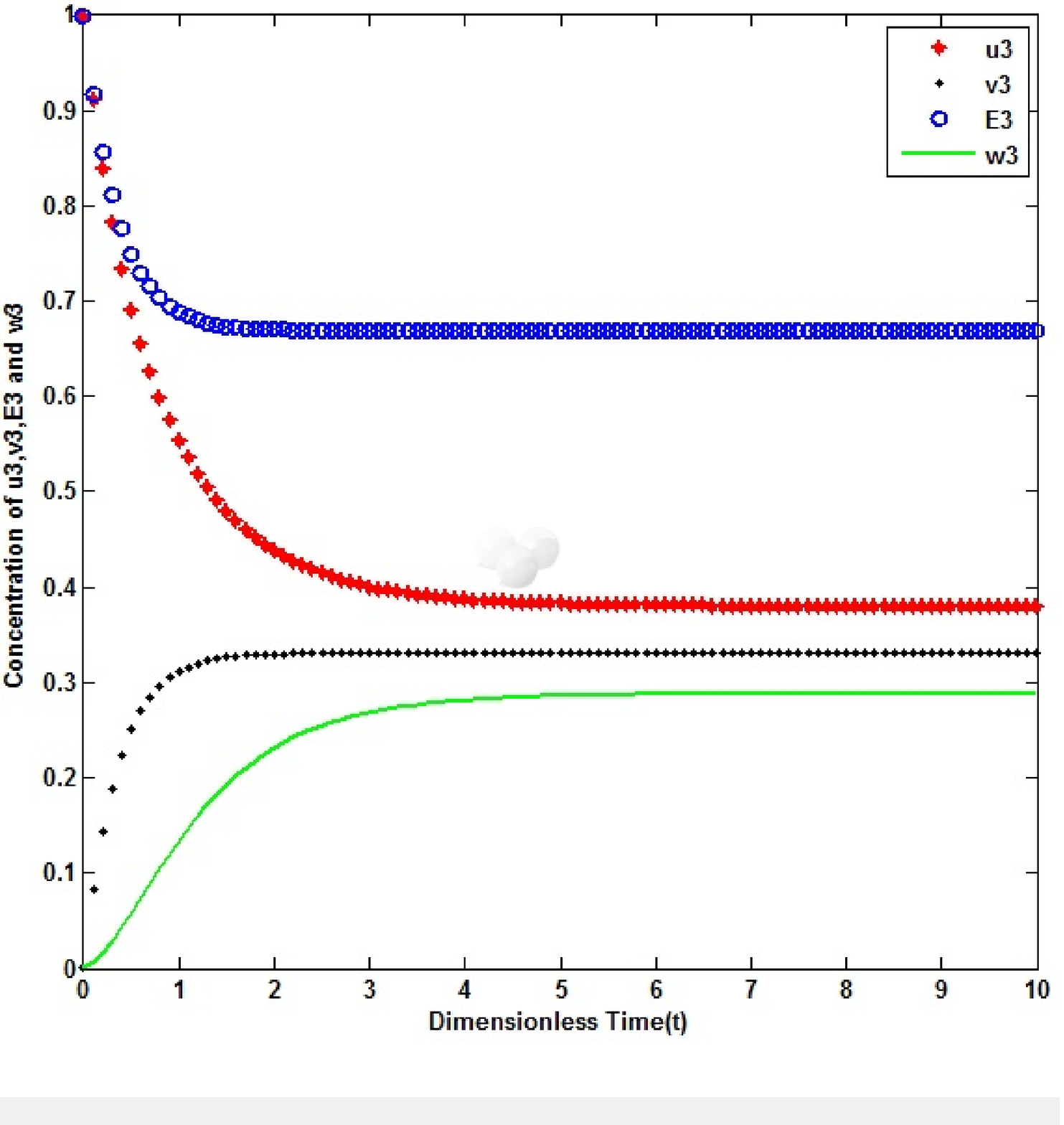}
\caption{$\varepsilon = 1.001, \alpha = 1.001, \lambda = 0.4$ and $k = 1.3$}
\label{fig:fig3a}
\end{figure}

\begin{figure}[h]
\center
\includegraphics[width=0.5\textwidth]{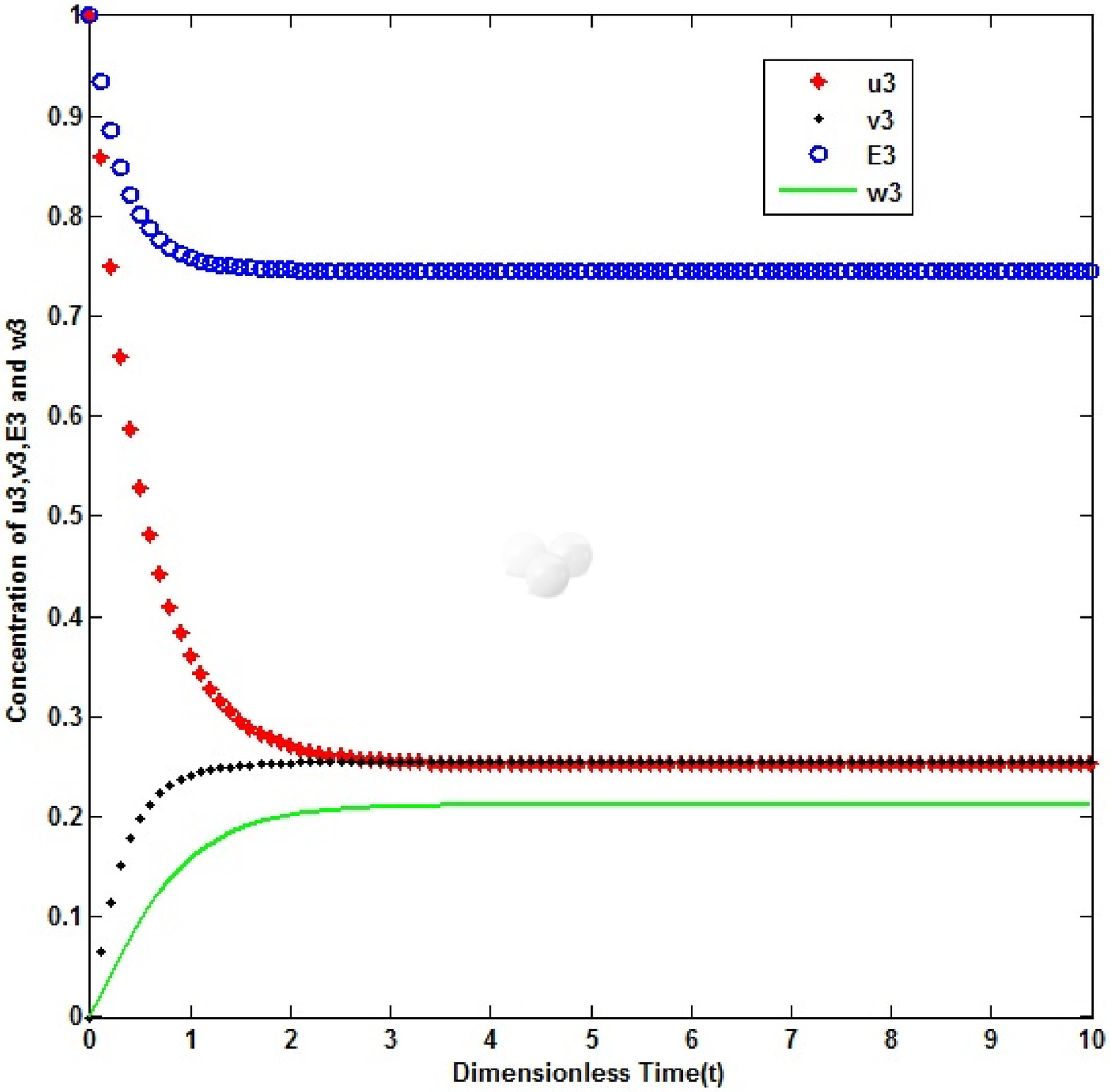}
\caption{$\varepsilon = 1.6, \alpha = 1.3, \lambda = 0.9$ and $k = 1.7$}
\label{fig:fig3b}
\end{figure}

\begin{figure}[h]
\center
\includegraphics[width=0.5\textwidth]{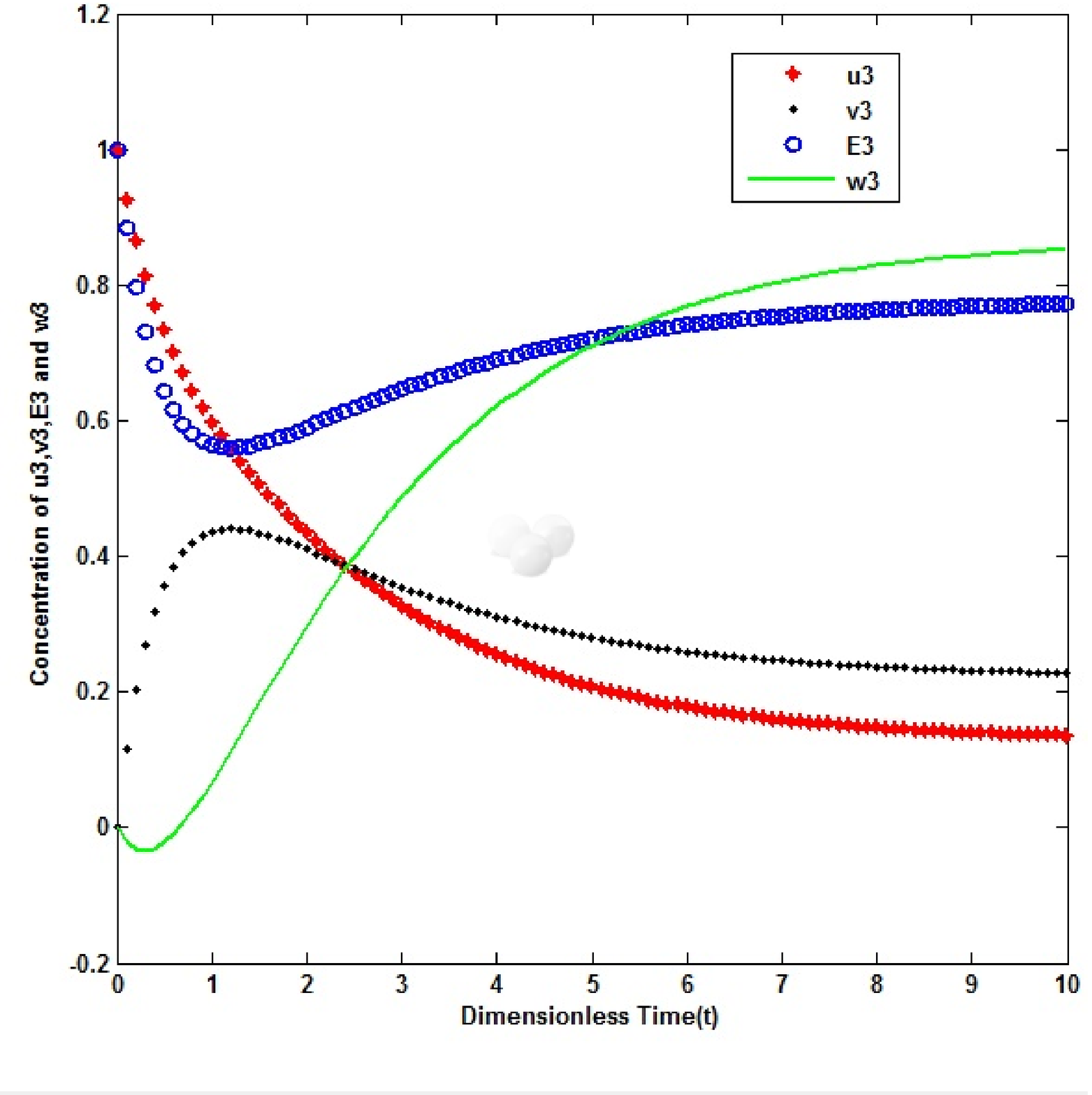}
\caption{$\varepsilon = 0.8, \alpha = 0.2, \lambda = 0.6$ and $k = 1.1$}
\label{fig:fig3c}
\end{figure}

\begin{figure}[h]
\center
\includegraphics[width=0.5\textwidth]{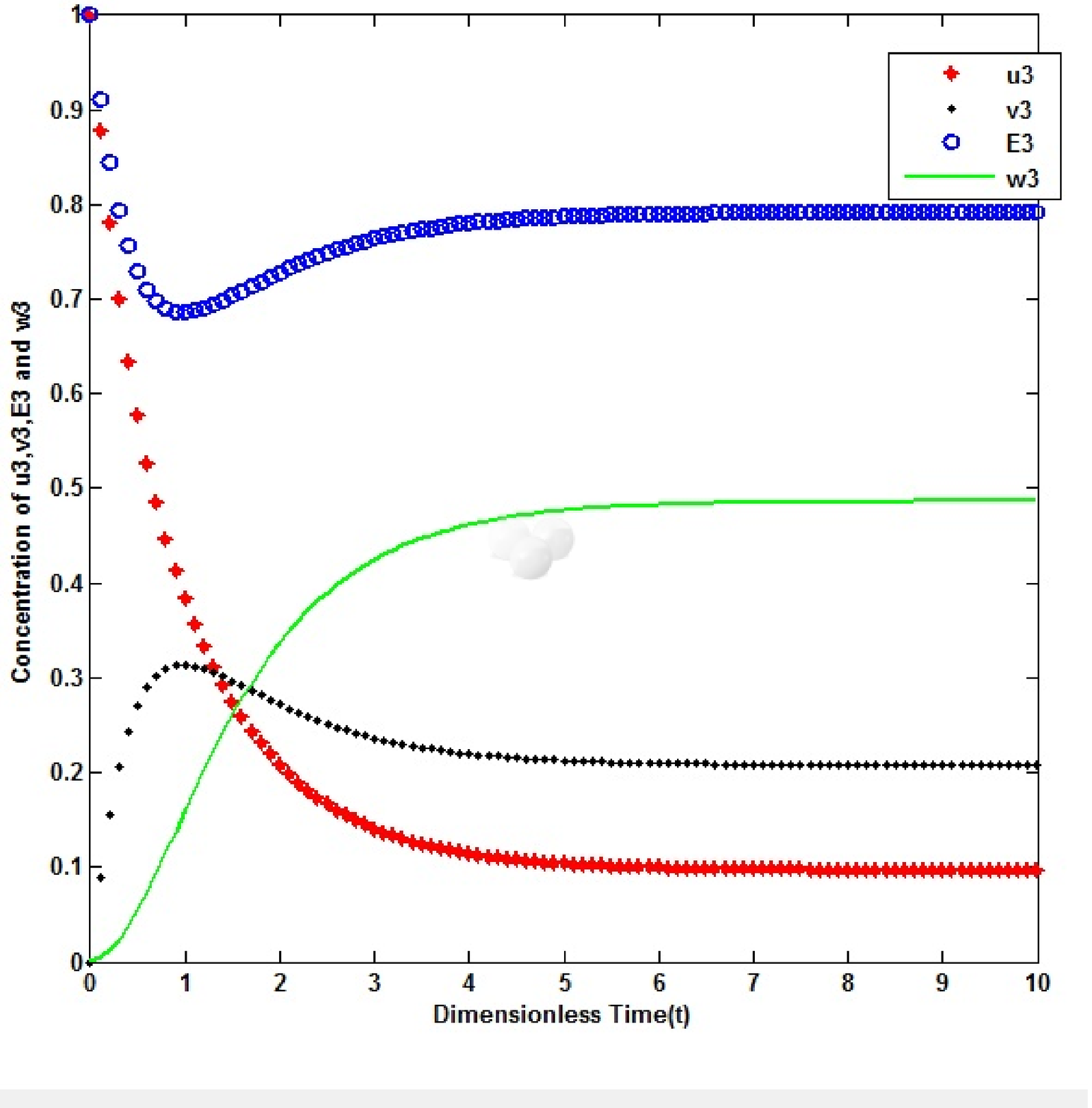}
\caption{$\varepsilon = 1.3, \alpha = 0.3, \lambda = 0.8$ and $k = 1.2$}
\label{fig:fig3d}
\end{figure}

\begin{figure}[h]
\center
\includegraphics[width=0.52\textwidth]{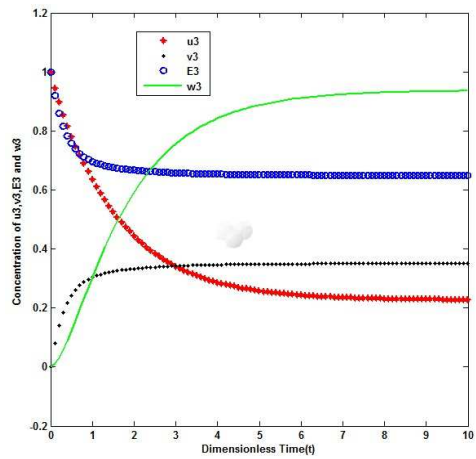}
\caption{$\varepsilon = 0.6, \alpha = 1.2, \lambda = 1.2$ and $k = 1.7$}
\label{fig:fig3e} 
\end{figure}

\begin{figure}[h]
\center 
\includegraphics[width=0.52\textwidth]{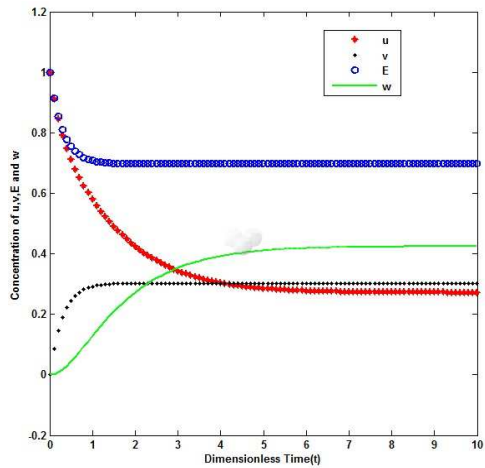}
\caption{$\varepsilon = 1, \alpha = 1, \lambda = 0.4$ and $k = 1.3$}
\label{fig:fig4a}
\end{figure}

\begin{figure}[h]
\center
\includegraphics[width=0.57\textwidth]{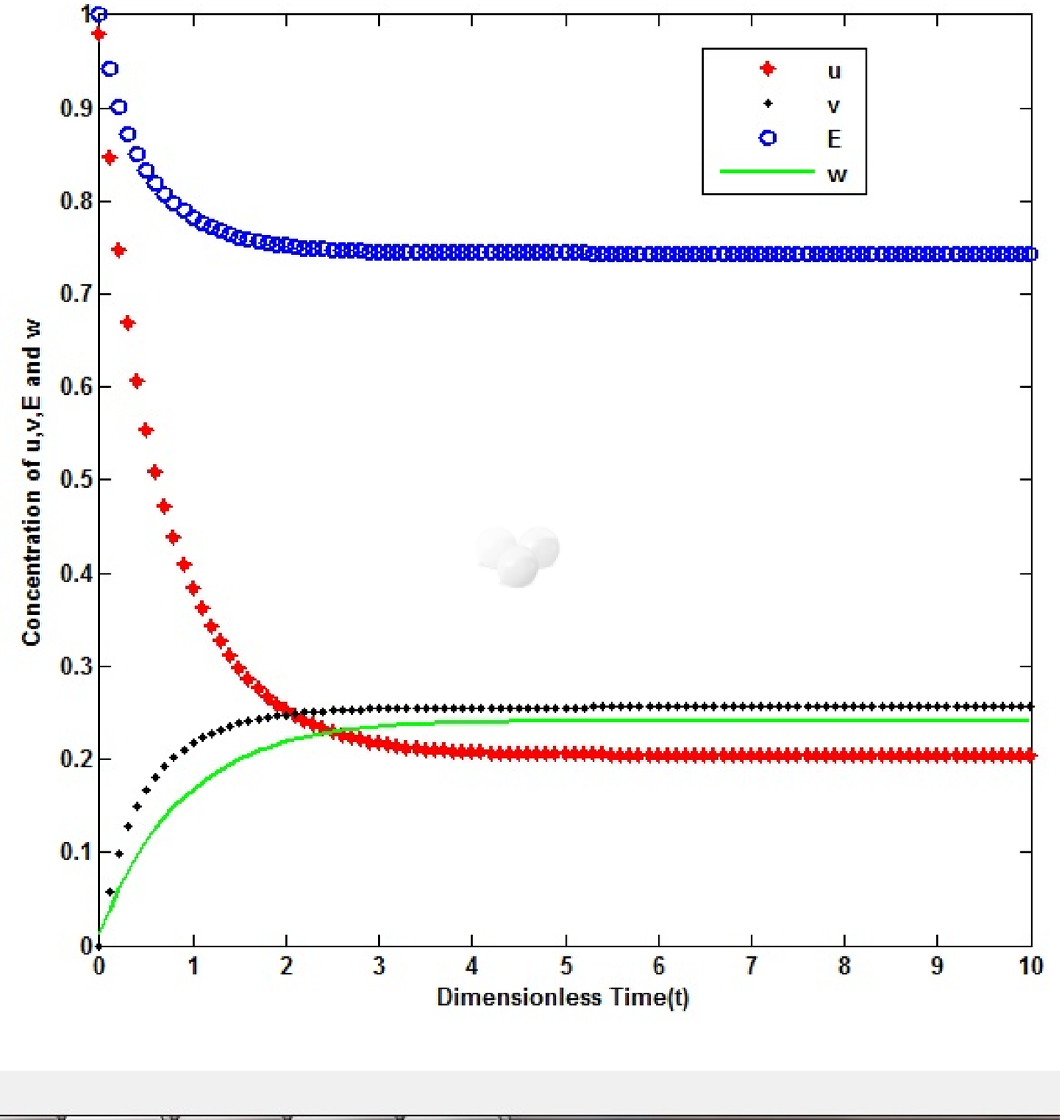}
\caption{$\varepsilon = 1.6, \alpha = 1.3, \lambda = 0.9$ and $k = 1.7$}
\label{fig:fig4b}
\end{figure}

\begin{figure}[h]
\center
\includegraphics[width=0.5\textwidth]{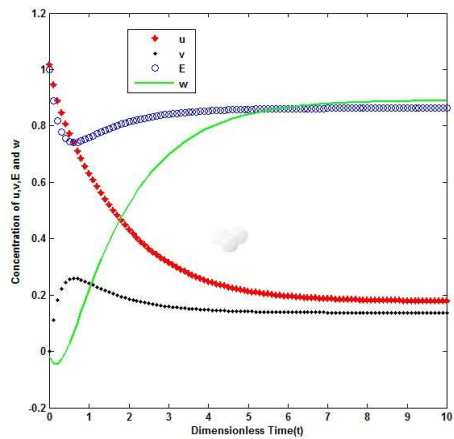}
\caption{$\varepsilon = 0.8, \alpha = 0.6, \lambda = 2$ and $k = 3.3$}
\label{fig:fig4c}
\end{figure}

\begin{figure}[h] 
\center
\includegraphics[width=0.5\textwidth]{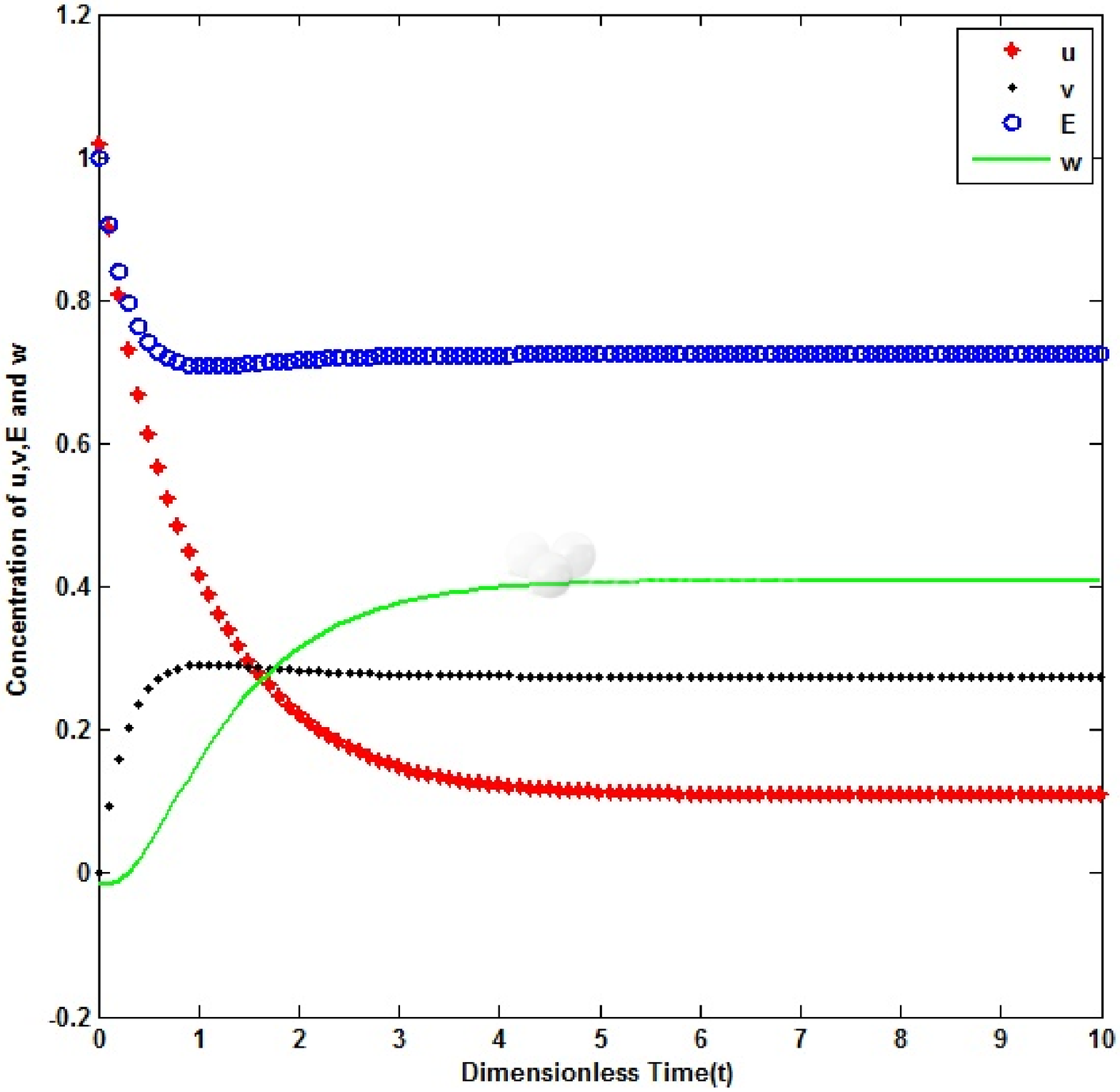}
\caption{$\varepsilon = 1.3, \alpha = 0.9, \lambda = 0.8$ and $k = 1.2$}
\label{fig:fig4d}
\end{figure}

\begin{figure}[h]
\center
\includegraphics[width=0.5\textwidth]{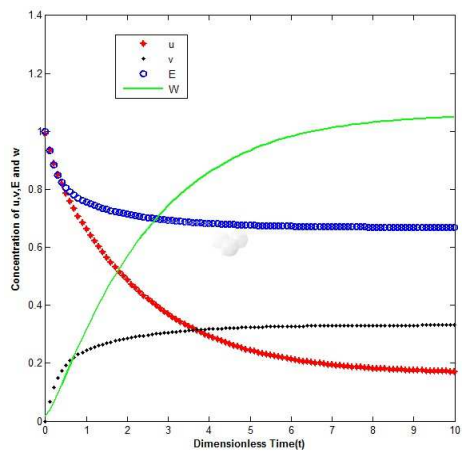}
\caption{$\varepsilon = 0.6, \alpha = 1.2, \lambda = 1.2$ and $k = 1.7$}
\label{fig:fig4e} 
\end{figure}

The figures in this section are divided in to four groups. The first three groups are related to three iterations of SIM and the last group refers to the HPM. Figures \ref{fig:fig1a}-\ref{fig:fig4e} show the analytic approximate solution of substrate $u$, enzyme $E$, enzyme-substrate complex $v$ and product $w$. Each figure in this work corresponds to one case in the previous section. The figures change in terms of the values of the dimensionless parameters $\varepsilon$, $\alpha$, $\lambda$ and $k$. We have applied two different methods which are SIM and HPM to find the analytical approximate solutions for equations (\ref{equ:eq18})-(\ref{equ:eq19}). The HPM has been used by many researchers for system (\ref{equ:eq1}) \cite{Varadharajan2011DD, Varadharajan2011D, Varadharajan2011, Maheswari2011}. The main purpose of this discussion is to find out the similarities and differences between the methods which are used in this study.  Another purpose is to recognize the best iteration of the SIM compared to the HPM.

\noindent There are a variety of data results that tell us the second iteration in our approach (SIM) is similar to  HPM. First of all, the second iteration has many significant similarities compared to (HPM), and some of them   provide excellent results in terms of our work. For instance, Figures \ref{fig:fig2a}-\ref{fig:fig2e} show that the value of the concentration of substrate $u$ slightly decreases from its initial value $(u(0) = 1)$, and there are a few changes in the value of the concentration of the enzyme-substrate complex $v$. Generally, they reach some constant values after $\tau > 4$. Also, in Figures \ref{fig:fig4a}-\ref{fig:fig4e} it appears that the concentration of the components are somewhat similar to those of corresponding Figures \ref{fig:fig2a}-\ref{fig:fig2e}. Another example is that the value of the concentration of enzyme $E$ in both sets of figures is more or less is the same, especially in cases 1, 2 and 5.  
     
\noindent Another crucial point is that the value of concentration $v$ in Figure \ref{fig:fig3c} reaches a maximum when $0 < \tau < 2$. Also, in the same interval of time, the value of the concentration $v$ reaches a maximum in Figure \ref{fig:fig4c} as well. We can also realize that the value of the enzyme in both figures ends up at a minimum value when $0 < \tau < 2$. In addition, Figures \ref{fig:fig3a}-\ref{fig:fig3e} and Figures \ref{fig:fig4a}-\ref{fig:fig4e} show that there is a gradual decrease in the rate of substrate $u$ between $0 < \tau < 2$ which then levels off after $\tau > 4$. On the other hand, the concentration of the product $w$ slightly increases between $0 < \tau < 2$ in both set of figurs , and is likely to remain stable after $\tau > 4$.

\noindent However,  there are some differences between our simple technique (SIM) and the classical technique (HPM). For example, Figures \ref{fig:fig1a}-\ref{fig:fig1e} show that the value of the concentration of substrate $u$ slightly decreases from its initial value $(u(0)=1)$, and there are a few changes in the value of the concentration of the enzyme-substrate complex $v$. Generally, they become zero after $\tau > 5$. Meanwhile, in Figures \ref{fig:fig4a}-\ref{fig:fig4e} it appears that the concentration of the components do not fall to zero, but instead reach some constant values. Basically, it could be pointed out that the differences between them are small and can be therefore be ignored.

\noindent Overall, it can be said that the second and third iterations of SIM are appropriate for obtaining a good approximate solution for our case study. In particular, the results of the second iteration are more fitted to an approximate solution in comparison with the classical technique (HPM). However, although there are some different values in terms of results between HPM and the second iteration method, they are tiny.
 \newpage  
  \paragraph{Figures \ref{fig:fig1a}-\ref{fig:fig1e}}. In these profiles of the normalized concentrations of the substrate $u$, enzyme-substrate complex $v$, enzyme $E$ and product $w$ correspond to case 1-case 5, respectively. The equations of step 1 are applied to plot the Figures (see Appendix C).
\newpage
\newpage
 \paragraph{Figures \ref{fig:fig2a}-\ref{fig:fig2e}}. In these profiles of the normalized concentrations of the substrate $u,$ enzyme-substrate complex $v,$ enzyme $E$ and product $w$ correspond to case 1-case 5, respectively. The equations of step 2 are applied to plot the Figures (see Appendix D).
\newpage
\newpage
 \newpage
\paragraph{Figures \ref{fig:fig3a}-\ref{fig:fig3e}}. In these profiles of the normalized concentrations of the substrate $u$, enzyme-substrate complex $v$, enzyme $E$ and product $w$ correspond to case 1-case 5, respectively. The equations of step 3 are applied to plot the Figures (see Appendix E).
\newpage
\newpage
  \paragraph{Figures \ref{fig:fig4a}-\ref{fig:fig4e}}. In these profiles of the normalized concentrations of the substrate $u,$ enzyme-substrate complex $v,$ enzyme $E$ and product $w$ correspond to case 1-case 5, respectively. The equations (\ref{equ:eq31})-(\ref{equ:eq35}) are applied to plot the Figures (see Appendix A).
\newpage
\newpage
\newpage 
\section{Findings}
\begin{figure}[h] 
\center
\includegraphics[width=0.8 \textwidth]{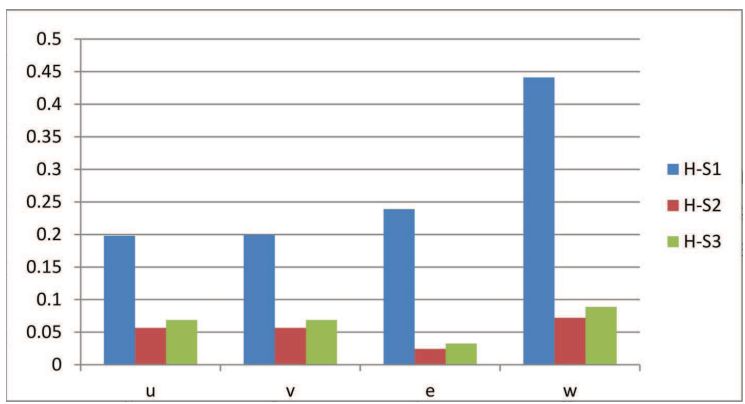}
\caption{The average value of the second norms convergence between the HPM and the iterations of the SIM. The blue column (H-S1), brown column (H-S2) and green column (H-S3) describe the second norm differences between (HPM) and the iterations of (SIM), respectively. The figure is plotted by using the results of Table \ref{tab:table1}-\ref{tab:table3}. The second norm differences are represented by  $u$, $v$, $e$ and $w$.}
\label{fig:vuw}
\end{figure}

We have used  the mean of the second norm (equation (\ref{equ:eq47})) to find the total differences between the HPM and each iteration of the SIM (see Tables \ref{tab:table1}-\ref{tab:table3} ). The rate of convergence between the SIM and the HPM is shown in Figure \ref{fig:vuw}. Thus, we use the following equation to find this rate of convergence:
  \begin{equation} \label{equ:eq47}
    \frac{\Vert f-g \Vert }{\sqrt{N}} = \sqrt{ \frac{\sum_{i=1}^{N} \vert f_{i} - g_{i} \vert ^{2} }{N}}
    \end{equation}
\noindent where $f$ and $g$ are the value of the concentrations of substances $u, v, E$ and $w$ for the SIM and the HPM respectively, and $N$ is the number of timescale iterations.  
 
The average norm between the second iteration and HPM is small in value. For instance, the average value of the norm concentration of $E$ is small (equal to 0.02) (see H-S2 in \ref{fig:vuw}). This means that the second iteration method is the most appropriate iteration in this case study in terms of approaching the approximate solution.   Although the rate of the second norm for the third iteration is also small (see H-S3), but the second iteration method of our work is the best iteration in order to obtain the convergence in terms of the solution in comparison with the classical method (HPM).

\noindent On the other hand, the differences between our approach (SIM) and the HPM occurred more frequently in the first iteration than in other iterations (see H-S1 values). It could be said that this iteration is not quite appropriate in this case study. This may be caused by giving the non-linear part in this iteration a zero value (see step 1).
  
  \begin{table} 
  \center
  \begin{tabular}{ | p{1.7cm} | l | l | l | l | l |  l |  }
    \hline
    Norm Con. & Case 1 & Case 2 & Case 3 & Case 4 & Case 5 & Average 1  \\ \hline
    UU1& 0.26747 &  0.20151&    0.17482&    0.16551&    0.18084&    0.19803 \\ \hline
    VV1 & 0.26918&  0.20313&    0.17535&    0.16627&    0.18385&    0.199556     \\ \hline
    EE1 &0.29684    & 0.24889 & 0.14721 &   0.18732 &   0.31446 &   0.238944    \\  \hline
    WW1 & 0.5633 &  0.37454 & 0.36304 &     0.29118 &   0.614 & 0.441212  \\ \hline
    \end{tabular}
    \caption{The average number of second norms between the first iteration (SIM) and HPM. The results are calculated by using Matlab program (see Appendices B and C).}
\label{tab:table1} 
\end{table}
\begin{table} 
 \center 
    \begin{tabular}{ | p{1.6cm} | l | l | l | l | l |  l |  }
    \hline
    Nor. Con. & Case 1 & Case 2 & Case 3 & Case 4 & Case 5 & Aver. 2  \\ \hline
    UU2& 0.03611 &  0.02504&    0.02326&    0.14113&    0.05728&    0.056564 \\ \hline
    VV2 & 0.03611 & 0.02507  & 0.02333 &    0.14114 &   0.05728 &   0.056586    \\ \hline
    EE2 & 2.204 E-17    & 0.01363&  0.02554 &   0.06548 &   0.01727 &   0.024384 \\  \hline
    WW2 & 0.03611    & 0.02008 &    0.05259 &   0.16332 &   0.08771 &   0.071962 \\ \hline
    \end{tabular}
    \caption{The average number of second norms between the second iteration (SIM) and HPM. The results are calculated by using Matlab program (see Appendices B and D).}
 \label{tab:table2}
\end{table}
\begin{table} 
\center
    \begin{tabular}{ | p{1.7cm} | l | l | l | l | l |  l |  }
    \hline
    Norm Con. & Case 1 & Case 2 & Case 3 & Case 4 & Case 5 & Average 3  \\ \hline
    UU3& 0.08435 &  0.04113 &   0.04704 & 0.13439 & 0.03628     & 0.068638 \\ \hline
    VV3 & 0.08439   & 0.04113 & 0.0471 &     0.13448     & 0.03633 &    0.068686 \\ \hline
    EE3 &0.02625 &  0.00942  & 0.02733 &    0.0672 &     0.03269 &  0.03257 \\  \hline
    WW3 & 0.10839 & 0.02597&    0.08244 &   0.15919 &   0.06823 &   0.088844 \\ \hline
    \end{tabular} 
    \caption{The average number of second norms between the third iteration (SIM) and HPM. The results are calculated by using Matlab program (see Appendices B and E).}
 \label{tab:table3}
\end{table}
 
\section{Conclusion}
 The simple iteration method (SIM) and the Homotopy Perturbation Method (HPM) are used to find approximate analytic solutions to non-linear differential equations (\ref{equ:eq18})-(\ref{equ:eq19}). Straightforward methods are derived for estimating the concentrations of substrate $u,$ product $w,$ enzyme-substrate complex $v$ and enzyme $E.$ The dimensionless technique applies to reduce the non-linear system of ODE.  
 
\noindent The HPM was used for a simple enzyme reaction (\ref{equ:eq1}) \cite{Maheswari2011, Varadharajan2011}. We have used this method for our case study, and have obtained an analytical approximate solution. Furthermore, a simple approach technique (SIM) was applied. This consisted of three iterations (steps). The approximate solution of the second step is similar to the classical method (HPM) (see Figures \ref{fig:fig2a}-\ref{fig:fig2e} and Figures \ref{fig:fig4a}-\ref{fig:fig4e}).

\noindent We have also used the idea of the second norm to determine the best iteration for the problem. So, it is clear that the second iteration method is quite similar to the HPM. Consequently, Figure \ref{fig:vuw} shows that the second iteration is the appropriate one (see Figure \ref{fig:vuw} for the H-S2 values). Thus, the SIM technique could be applied to some other complex chemical reactions to find appropriate solutions, and to describe the behaviour of their parameters. For example, it could be applied to many open path ways in terms of biochemical reactions \cite{Flach2010}. In addition, we highly recommend applying the simple approach (SIM) to describe the approximate solutions of complex enzyme reactions \cite{Pedersen2008}, the reaction mechanism of competitive inhibitions, and the reaction scheme of allosteric inhibitions \cite{Goeke2012}.

  \paragraph{Appendix A} 
This appendix consists of the solution of equations (\ref{equ:eq18}) and (\ref{equ:eq19})) by using the HPM. Furthermore, this method is used to derive equations (\ref{equ:eq31}) and (\ref{equ:eq32}) from equations (\ref{equ:eq18}) and (\ref{equ:eq19}), let $a = \varepsilon(k-\lambda), b = 1-\alpha$ and  $c = k+\alpha+\alpha \varepsilon$,
 \begin{equation} \label{equ:eq48}
  (1-q) [\frac{du}{d\tau}+\varepsilon u]+q[\frac{du}{d\tau}+\varepsilon u-av-\varepsilon v] = 0 ,
  \end{equation}
  \begin{equation}\label{equ:eq49}
  (1-q) [\frac{dv}{d\tau} +cv]+q[\frac{dv}{d\tau}-bu+cv+buv-\alpha \varepsilon v^{2} -\alpha] = 0,
  \end{equation}
\noindent with initial conditions,
  \begin{equation}\label{equ:eq50}
  u(0) = 1, v(0) = 0
  \end{equation}
\noindent Thus, by using the HPM \cite{Maheswari2011, Varadharajan2011, Varadharajan2011D}, the approximate solution of equations (\ref{equ:eq48}) and (\ref{equ:eq49}) are:
  \begin{equation}\label{equ:eq51}
  u = u_{0}+qu_{1}+q^{2} u_{2}+...
  \end{equation}
  \begin{equation}\label{equ:eq52}
   v = v_{0}+qv_{1}+q^{2} v_{2}+...
  \end{equation}
\noindent Substituting the equations (\ref{equ:eq51}) and (\ref{equ:eq52}) in equations (\ref{equ:eq48}) and (\ref{equ:eq49}), and comparing the coefficients of the like power $q,$ we can get the following system of ordinary differential equations:
 \begin{equation} \label{equ:eq53}
  q^{0}: \frac{du_{0}}{d\tau}+\varepsilon u_{0} = 0
 \end{equation}
  \begin{equation} \label{equ:eq54}
  q^{1}: \frac{du_{1}}{d\tau}+\varepsilon u_{1}-a v_{0}-\varepsilon u_{0} v_{0} = 0
  \end{equation}
  \begin{equation} \label{equ:eq55}
  q^{2}: \frac{du_{2}}{d\tau}+ \varepsilon u_{2}- av_{1} -\varepsilon u_{0} v_{1}-\varepsilon u_{1} v_{0} = 0
  \end{equation}
\noindent and,
  \begin{equation} \label{equ:eq56}
  q^{0}: \frac{dv_{0}}{d \tau}+c v_{0} = 0
  \end{equation}
  \begin{equation}\label{equ:eq57}
  q^{1}: \frac{dv_{1}}{d\tau}+c v_{1}- bu_{0}+b u_{0} v_{0}- \alpha \varepsilon v_{0} ^{2} -\alpha = 0,
  \end{equation}
\begin{equation} \label{equ:eq58}
  q^{2}: \frac{dv_{2}}{d\tau}+c v_{2} -bu_{1}+b u_{0} v_{1}+bu_{1} v_{0}-2 \alpha \vartheta v_{0} v_{1} = 0,
\end{equation}
\noindent By solving the equations ( \ref{equ:eq53})-( \ref{equ:eq58}) using initial conditions equation (\ref{equ:eq50}), we obtain the following solutions:
  \begin{equation}\label{equ:eq59}
  u_{0}(\tau) = e^{-\varepsilon \tau}
  \end{equation}
  \begin{equation}\label{equ:eq60}
  u_{1}(\tau) = e^{-\varepsilon \tau}
  \end{equation}
 \begin{equation}\label{equ:eq61}
 \begin{split}
u_{2}(\tau) = & (\frac{ab}{c-\varepsilon}+\frac{\alpha \varepsilon}{c}) t e^{-\varepsilon \tau} + \frac{abc-a \alpha \varepsilon+a \alpha c}{c (\varepsilon-c)^{2}} e^{-c \tau} \\&+ \frac{\alpha \varepsilon ^{2} -cb \varepsilon -c\alpha \varepsilon}{c^{2}(\varepsilon-c)} e^{(-\varepsilon-c) \tau}+  \frac{a \alpha}{c\varepsilon} +\frac{b}{\varepsilon-c} e^{-2 \varepsilon \tau} \\& + (-1+\frac{a\alpha \varepsilon - abc -ac \alpha}{c (\varepsilon-c)^{2}} - \frac{a \alpha}{c \varepsilon}+\frac{b}{c-\varepsilon}+\frac{-\alpha \varepsilon^{2} + cb \varepsilon +c \alpha \varepsilon }{c^{2} (\varepsilon -c)} ) e^{- \varepsilon \tau}
\end{split}
 \end{equation}
 
\noindent and,

 \begin{equation}\label{equ:eq62}
 v_{0}(\tau) = 0
 \end{equation}
 \begin{equation}\label{equ:eq63}
  v_{1}(\tau) = \frac{b}{c-\varepsilon} e^{-\varepsilon \tau} + \frac{bc-\alpha \varepsilon +c \alpha}{c(\varepsilon-c)} e^{-c \tau}
  \end{equation}
  \begin{equation}\label{equ:eq64}
  \begin{split}
v_{2}(\tau) = & (\frac{b}{c-\varepsilon}+\frac{b \alpha}{c(c-\varepsilon)}) e^{-\varepsilon \tau} + \frac{b^{2}}{(c-\varepsilon)(c-2 \varepsilon)} e^{-2 \varepsilon \tau} \\&+ \frac{\alpha \varepsilon b - c b^{2} -cb \alpha}{c \varepsilon (\varepsilon -c)} e^{(-\varepsilon -c) \tau} \\&+ ( \frac{b}{\varepsilon -c} + \frac{b^{2}}{(\varepsilon -c)(c-2 \varepsilon)}+ \frac{c b^{2}-b \alpha \varepsilon + cb \alpha}{c \varepsilon(\varepsilon -c)}+ \frac{\alpha b}{c(\varepsilon -c)}) e^{-c \tau} +\frac{\alpha}{c }
\end{split} 
  \end{equation}
\noindent According to the HPM, we can easily find that
  \begin{equation}\label{equ:eq65}
  u(\tau) = \lim_{q \to 1} u(\tau) = u_{0}+u_{1}+u_{2}+...
  \end{equation}
\noindent and,
  \begin{equation}\label{equ:eq66}
   v(\tau) = \lim_{q \to 1} v(\tau) = v_{0}+v_{1}+v_{2}+..
  \end{equation}
\noindent By putting equations (\ref{equ:eq59})-(\ref{equ:eq61}) in equation (\ref{equ:eq65}) and equations (\ref{equ:eq62})-(\ref{equ:eq64}) in equation (\ref{equ:eq66}), we obtain the approximate solution for a system of non-linear ODE equations (\ref{equ:eq18})-(\ref{equ:eq19}) which is described in equations (\ref{equ:eq31}) and (\ref{equ:eq32}).
\paragraph{Appendix B} 
 
\noindent Let $k1 =  \varepsilon $, $k2 =  \lambda, $ $k3 = \alpha $ and $t = \tau,$ and we use the following Matlab programming to plot the functions of the equations (\ref{equ:eq31})-(\ref{equ:eq35}).

\noindent t = 0;

\noindent for i = 1:101

\noindent k1 = 1 ; k2 = 1.2; k3 = 0.9; k = 1.3;

\noindent a = k1*(k-k2);  b = 1-k3; c = k+k3*k1+k3;

\noindent u =  2*exp(-k1*t)+((a*b)/(c-k1)+(k3*k1)/c)*t*exp(-k1*t)

\noindent + (a*b*c-a*k3*k1+a*k3*c)/(c*$(k1 -c)^2$)*exp(-c*t)

\noindent +  b/(k1-c)* exp(-2*k1*t)+(k3*$k1^2$-c*b*k1-k3*c*k1)/($c^2$*(k1-c))

\noindent *exp((-k1-c)*t)+  (a*k3)/(c*k1)

\noindent +(-1+(a*k1*k3-a*b*c -a*k3*c)/(c*$(k1-c)^2$)-(a*k3)/(c*k1)

\noindent + b/(c-k1)+(a*b*k1-k3*$k1^2$+k3*c*k1)/($c^2$*(k1-c)))*exp(-k1*t);

\noindent v = b/(c-k1)*exp(-k1*t)+((c*b-k3*k1+k3*c)/(c*(k1-c)))*exp(-c*t)+k3/c

\noindent + (b/(c-k1)+(k3*b)/(c*(c-k1)))*exp(-k1*t)

\noindent +($b^2$/((c-k1)*(c-2*k1)))*exp(-2*k1*t)

\noindent +(k3*k1*b-c*$b^2$-k3*c*b)/(c*k1*(k1-c))*exp((-c-k1)*t)

\noindent +(b/(k1-c)+$b^2$/((k1-c)*(c-2*k1))+(c*$b^2$-k3*k1*b

\noindent +k3*c*b)/(c*k1*(k1-c))+(k3*b)/(c*(k1-c)))*exp(-c*t);

\noindent E = 1-v; w = 1/k1 -u/k1 -v; A(i) = u; B(i) = v ; C(i) = E; D(i) = w; T(i) = t; 

\noindent t = t+0.1; 

\noindent end 

\noindent plot(T,A,'r',T,B,'k.',T,C,'b.',T,D,'g')  

\noindent ylabel('Concentration of $u, v, E$ and $w$') 

\noindent xlabel('Dimensionless Time($t$)') 

\noindent axis square 
\paragraph{Appendix C} 
  
\noindent Let $k1 = \varepsilon,$ $k2 =  \lambda$,  $k3 = \alpha$ and $t = \tau,$ and we use the following Matlab programming to plot the functions of step 1.

\noindent t = 0;

\noindent for i = 1:101;

\noindent k1 = 1;  k2 = 1.2; k3 = 0; k = 1.3; a=k1*(k-k2);  b=1-k3; c=k+k3+k3*k1; 

\noindent  p1 = (-k1-c-sqrt($(k1+c)^2$-4*(c*k1-a*b)))/2 ; p2 = (-k1-c+sqrt($(k1+c)^2$-4*(c*k1-a*b)))/2 ; 

\noindent d2 = (p2+k1)/(p2-p1); d3 = (p1+k1)/(p1-p2);

\noindent  d1 = ((p1+k1)*(p2+k1))/(a*(p2-p1)); 

\noindent  u1 = d2*exp(p1*t)+d3*exp(p2*t); v1 = d1*exp(p1*t)-d1*exp(p2*t); 

\noindent e1 = 1-v1; w1 = 1/k1 -u1/k1 -v1; 

\noindent A(i) = u1; B(i) = v1; C(i) = e1; D(i) = w1; T(i) = t; 

\noindent t = t+0.1; 

\noindent end 

\noindent \ plot(T,A,'r',T,B,'k.',T,C,'b.',T,D,'g') 

\noindent  ylabel('Concentration of $u1, v1, E1$ and $w1$') 

\noindent xlabel('Dimensionless Time($t$)') 

\noindent axis square 


\paragraph{Appendix D} 

\noindent Let $k1 = \varepsilon,$ $k2 = \lambda,$ $k3 = \alpha$ and $t = \tau,$ and we use the following Matlab programming to plot the functions of step 2.

\noindent t = 0;

\noindent for i = 1:101;

\noindent k1 = 1; k2 = 1.2; k3 = 0.2; k = 1.3; a = k1*(k-k2); b = 1-k3; c = k+k3+k3*k1; 

\noindent  p1 = (-k1-c-sqrt($(k1+c)^2$-4*(c*k1-a*b)))/2 ; 

\noindent p2 = (-k1-c+sqrt($(k1+c)^2$-4*(c*k1-a*b)))/2 ; 

\noindent d2 = (p2+k1)/(p2-p1); d3 = (p1+k1)/(p1-p2); 

\noindent d1 = ((p1+k1)*(p2+k1))/(a*(p2-p1)); 

\noindent d4 = k1*d2*d1; d5 = (-k1*d2*d1)+(k1*d3*d1); 

\noindent  d6 = -k1*d3*d1; d7 = (-b*d2*d1)+(k3*k1*$d1^2$ ; 

\noindent d8 = (b*d2*d1)-(b*d3*d1)-(2*k3*k1*$d1^2$); 

\noindent d9 = (b*d3*d1)+(k3*k1*$d1^2$) ; d10 = (p2+k1)/(a*(p2-p1)); 

\noindent d11 = -1/(p2-p1); d12 = (p1+k1)/(a*(p1-p2)); 

\noindent d13 = 1/(p2-p1); d14 = (d4*d10+d11*d7)/p1 ; 

\noindent d15 = (d10*d5+d11*d8)/p2 ; d16 = (d10*d6+d11*d9)/(2*p2-p1); 

\noindent d17 = (-k3*d11)/p1 ; d18 = (d12*d4+d13*d7)/(2*p1-p2); 

\noindent d19 = (d12*d5+d13*d8)/p1 ; d20 = (d12*d6+d13*d9)/p2 ; 

\noindent d21 = (-k3*d13)/p2 ; d22 = a*d14+a*d18; d23 = a*d15+a*d19; 

\noindent d24 = a*d16+a*d20; d25 = a*d17+a*d21; d26 = (p1+k1)*d14+(p2+k1)*d18; 

\noindent d27 = (p1+k1)*d15+(p2+k1)*d19 ; d28 = (p1+k1)*d16+(p2+k1)*d20 ; 

\noindent  d29 = (p1+k1)*d17+(p2+k1)*d21 ;  

\noindent c3 = (1-d22-d23-d24-d25)/a  -(a*(d26+d27+d28+d29)

\noindent +(p1+k1)*(1-d22-d23-d24-d25))/(a*(p1-p2)); 

\noindent c4 = (a*(d26+d27+d28+d29)

\noindent + (p1+k1)*(1-d22-d23-d24-d25))/(a*(p1-p2)); 

\noindent u2 = c3*a*exp(p1*t)+c4*a*exp(p2*t)+d22*exp(2*p1*t)

\noindent +d23*exp((p1+p2)*t)+d24*exp(2*p2*t)+d25; 

\noindent v2 = c3*(p1+k1)*exp(p1*t)+c4*(p2+k1)*exp(p2*t)

\noindent +d26*exp(2*p1*t)+d27*exp((p1+p2)*t)+d28*exp(2*p2*t)+d29; 

\noindent e2 = 1-v2; w2 = 1/k1 -u2/k1 -v2; 

\noindent A(i) = u2; B(i) = v2; C(i) = e2; D(i) = w2; T(i) = t;

\noindent t = t+0.1;

\noindent end 

\noindent plot(T,A,'r',T,B,'k.',T,C,'b.',T,D,'g') 

\noindent ylabel('Concentration of $u2, v2, E2$ and $w2$') 

\noindent xlabel('Dimensionless Time($t$)') 

\noindent axis square 


\paragraph{Appendix E} 
 
\noindent Let $k1 = \varepsilon,$ $k2 = \lambda,$ $k3 = \alpha$ and $t = \tau,$ and we use the following Matlab programming to plot the functions of step 3.

\noindent t = 0;  

\noindent for i = 1:101;

\noindent k1 = 1; k2 = 1.2; k3 = 0.2; k = 1.4; a = k1*(k-k2);

\noindent b = 1-k3; c=k+k3+k3*k1;  

\noindent p1 = (-k1-c-sqrt($(k1+c)^2$-4*(c*k1-a*b)))/2; 

\noindent p2 = (-k1-c+sqrt($(k1+c)^2$-4*(c*k1-a*b)))/2 ; 

\noindent d2 = (p2+k1)/(p2-p1); d3 = (p1+k1)/(p1-p2); 

\noindent d1 = ((p1+k1)*(p2+k1))/(a*(p2-p1)); d4 = k1*d2*d1;  

\noindent d5 = (-k1*d2*d1)+(k1*d3*d1);

\noindent d6 = -k1*d3*d1; d7 = (-b*d2*d1)+(k3*k1*$d1^2$); 

\noindent d8 = (b*d2*d1)-(b*d3*d1)-(2*k3*k1*$d1^2$);d9 = (b*d3*d1)+(k3*k1*$d1^2$); 

\noindent d10 = (p2+k1)/(a*(p2-p1)); d11=-1/(p2-p1); d12 = (p1+k1)/(a*(p1-p2)); 

\noindent d13 = 1/(p2-p1); d14=(d4*d10+d11*d7)/p1 ;d15 = (d10*d5+d11*d8)/p2 ; 

\noindent d16 = (d10*d6+d11*d9)/(2*p2-p1); d17 = (-k3*d11)/p1; 

\noindent d18 = (d12*d4+d13*d7)/(2*p1-p2); d19 = (d12*d5+d13*d8)/p1 ; 

\noindent d20 = (d12*d6+d13*d9)/p2 ; d21 = (-k3*d13)/p2 ; d22 = a*d14+a*d18; 

\noindent d23 = a*d15+a*d19; d24 = a*d16+a*d20; d25 = a*d17+a*d21;

\noindent d26 = (p1+k1)*d14+(p2+k1)*d18; d27 = (p1+k1)*d15+(p2+k1)*d19 ; 

\noindent d28 = (p1+k1)*d16+(p2+k1)*d20 ; d29 = (p1+k1)*d17+(p2+k1)*d21 ; 

\noindent c3 = (1-d22-d23-d24-d25)/a-(a*(d26+d27+d28+d29)

\noindent + (p1+k1)*(1-d22-d23-d24-d25))/(a*(p1-p2)); 

\noindent c4 = (a*(d26+d27+d28+d29)+(p1+k1)*(1-d22-d23-d24-d25))/(a*(p1-p2)); 

\noindent d30 = k1*$c3^2$*a*(p1+k1)+k1*d22*d29+k1*d26*d25; 

\noindent d31 = k1*c3*c4*a*(p2+k1)+k1*c4*c3*a*(p1+k1)+k1*d23*d29+k1*d25*d27;

\noindent d32 = k1*c3*a*d26+k1*c3*(p1+k1)*d22; 

\noindent d33 = k1*c3*a*d27+k1*c4*a*d26+k1*c4*(p2+k1)*d22+k1*c3*(p1+k1)*d23;

\noindent d34 = k1*c3*a*d28+k1*c4*a*d27+k1*c4*(p2+k1)*d23+k1*c3*(p1+k1)*d24; 

\noindent d35 = k1*c3*a*d29+k1*c3*(p1+k1)*d25; 

\noindent d36 = k1*$c4^2$ *a*(p2+k1)+k1*d24*d29+k1*d25*d28;

\noindent d37 = k1*c4*a*d28+k1*c4*(p2+k1*d24; 

\noindent d38 = k1*c4*a*d29+k1*c4*(p2+k1)*d2;

\noindent d39 = k1*d22*d26; d40 = k1*d22*d27+k1*d23*d26; 

\noindent  d41 = k1*d22*d28+k1*d23*d27+k1*d24*d26; d42 = k1*d23*d28+k1*d24*d27;

\noindent d43 = k1*d24*d28; d44 = k1*d25*d29; 

\noindent d45 = -b*d30/k1+k3*k1*$c3^2$*$(p1+k1)^2$+k3*k1*d26*d29+k3*k1*d29*d26;

\noindent d46 = -b*d31/k1+k3*k1*c3*c4*(p1+k1)*(p2+k1)

\noindent + k3*k1*c3*c4*(p1+k1)*(p2+k1)+k3*k1*d27*d29+k3*k1*d27*d29;

\noindent d47 = -b*d32/k1+k3*k1*c3*d26*(p1+k1)+k3*k1*c3*d26*(p1+k1); 

\noindent d48 = -b*d33/k1+k3*k1*c3*d27*(p1+k1)+k3*k1*c4*d26*(p2+k1); 

\noindent d49 = -b*d34/k1+k3*k1*c3*d28*(p1+k1)+k3*k1*c4*d27*(p2+k1)

\noindent + k3*k1*c4*d27*(p2+k1)+k3*k1*c3*d28*(p1*k1); 

\noindent d50 = -b*d35/k1+k3*k1*c3*d29*(p1+k1)+k3*k1*c3*d29*(p1*k1); 

\noindent d51 = -b*d36/k1+k3*k1*$c4^2$*(p2+k1)+k3*k1*d28*d29+k3*k1*d28*d29; 

\noindent d52 = -b*d37/k1+k3*k1*c4*d28*(p2+k1)+k3*k1*c4*d28*(p2+k1); 

\noindent d53 = -b*d38/k1+k3*k1*c4*d29*(p2+k1)+k3*k1*c4*d29*(p2+k1); 

\noindent d54 = -b*d39/k1+k3*k1*$(d26)^2$; 

\noindent d55 = -b*d40/k1+k3*k1*d26*d27+k3*k1*d26*d27; 

\noindent d56 = -b*d41/k1+k3*k1*d26*d28+k3*k1*$(d27)^2$+k3*k1*d26*d28; 

\noindent d57 = -b*d42/k1+k3*k1*d27*d28+k3*k1*d27*d28; 

\noindent d58 = -b*d43/k1+k3*k1*$(d28)^2$; d59 = -b*d44/k1+$(d29)^2$+k3; 

 \noindent d60 = (d10*d30+d11*d45)/(p1); d61 = (d10*d31+d11*d46)/(p2); 

\noindent d62 = (d10*d32+d11*d47)/(2*p1); d63 = (d10*d33+d11*d48)/(p1+p2); 

\noindent d64 = (d10*d34+d11*d49)/(2*p2); d65 = (d10*d35+d11*d50);

\noindent d66 = (d10*d36+d11*d51)/(-p1+2*p2); 

\noindent d67 = (d10*d37+d11*d52)/(-p1+3*p2);

\noindent d68 = (d10*d38+d11*d53)/(-p1+p2); d69 = (d10*d39+d11*d54)/(3*p1);

\noindent d70 = (d10*d40+d11*d55)/(2*p1+p2); d71 = (d10*d41+d11*d56)/(p1+2*p2);

\noindent d72 = (d10*d42+d11*d57)/(3*p2); d73 = (d10*d43+d11*d58)/(-p1+4*p2);

\noindent d74 = (d10*d44+d11*d59)/(-p1); d75 = (d12*d30+d13*d45)/(2*p1-p2);

\noindent d76 = (d12*d31+d13*d46)/(p1); d77 = (d12*d32+d13*d47)/(3*p1-p2);

\noindent d78 = (d12*d33+d13*d48)/(2*p1); d79 = (d12*d34+d13*d49)/(p1+p2);

\noindent d80 = (d12*d35+d13*d50)/(p1-p2); d81 = (d12*d36+d13*d51)/(p2);

\noindent d82 = (d12*d37+d13*d52)/(2*p2);  d83 = (d12*d38+d13*d53);

\noindent d84 = (d12*d39+d13*d54)/(4*p1-p2); d85 = (d12*d40+d13*d55)/(3*p1);

\noindent d86 = (d12*d41+d13*d56)/(2*p1+p2); d87 = (d12*d42+d13*d57)/(p1+2*p2);

\noindent d88 = (d12*d43+d13*d58)/(3*p2); d89 = (d12*d44+d13*d59)/(-p2);

\noindent d90 = a*d60+a*d75;d91=a*d61+a*d76; 

\noindent d92 = a*d62+a*d77; d93=a*d63+a*d78;

\noindent d94 = a*d64+a*d79; d95 = a*d65; d96 = a*d80; d97 = a*d66+a*d81;

\noindent d98 = a*d67+a*d82; d99 = a*d68; d100 = a*d83; d101 = a*d69+a*d84;

\noindent d102 = a*d70+a*d85; d103 = a*d71+a*d86; d104 = a*d72+a*d87; 

\noindent d105 = a*d73+a*d88; d106 = a*d74+a*d89; h1 = p1+k1; h2 = p2+k1; 

\noindent d107 = h1*d60+h2*d75; d108 = h1*d61+h2*d76; d109 = h1*d62+h2*d77;

\noindent d110 = h1*d63+h2*d78; d111 = h1*d64+h2*d79; 

\noindent d112 = h1*d65; d113 = h2*d80;

\noindent d114 = h1*d66+h2*d81; d115 = h1*d67+h2*d82; 

\noindent d116 = h1*d68; d117 = h2*d83;

\noindent d118 = h1*d69+h2*d84; d119 = h1*d70+h2*d85; d120 = h1*d71+h2*d86; 

\noindent d121 = h1*d72+h2*d87; d122 = h1*d73+h2*d88; d123 = h1*d74+h2*d89; 

\noindent M = d90+d91+d92+d93+d94+d96+d97+d98+d99

\noindent +d101+d102+d103+d104+d105+d106; 

\noindent N = d107+d108+d109+d110+d111+d113+d114+d115

\noindent +d116+d118+d119+d120+d121+d122+d123; 

\noindent c5 = ((1-M)*h2+a*N)/(a*(h2-h1)); c6=-N/h2-(h1/h2)*c5;

\noindent u3 = c5*a*exp(p1*t)+c6*a*exp(p2*t)+d90*exp((2*p1)*t)

\noindent +d91*exp((p1+p2)*t)+d92*exp((3*p1)*t)

 \noindent +d93*exp((2*p1+p2)*t)+d94*exp((p1+2*p2)*t)+d95*t*exp((p1)*t)

 \noindent +d96*exp((p1)*t)+d97*exp((2*p2)*t)+d98*exp((3*p2)*t)

 \noindent +d99*exp((2*p2)*t)+d100*t*exp((2*p2)*t)+d101*exp((4*p1)*t)

  \noindent +d102*exp((3*p1+p2)*t)+d103*exp((2*p1+2*p2)*t)

 \noindent +d104*exp((p1+3*p2)*t)+d105*exp((4*p2)*t)+d106; 

\noindent v3 = c5*h1*exp(p1*t)+c6*h2*exp(p2*t)+d107*exp((2*p1)*t)

\noindent +d108*exp((p1+p2)*t)+d109*exp((3*p1)*t)

\noindent +d110*exp((2*p1+p2)*t)+d111*exp((p1+2*p2)*t)+d112*t*exp((p1)*t)

\noindent +d113*exp((p1)*t)+d114*exp((2*p2)*t)+d115*exp((3*p2)*t)

\noindent +d116*exp((2*p2)*t)+d117*t*exp((2*p2)*t)

\noindent +d118*exp((4*p1)*t)+d119*exp((3*p1+p2)*t)+d120*exp((2*p1+2*p2)*t)

\noindent +d121*exp((p1+3*p2)*t)+d122*exp((4*p2)*t)+d123;

\noindent e3 = 1-v3; w3 = 1/k1 -u3/k1 -v3; 

\noindent A(i) = u3; B(i) = v3; C(i) = e3; D(i) = w3; T(i) = t; 

\noindent t = t+0.1; 

\noindent end  

\noindent plot(T,A,'r',T,B,'k.',T,C,'b.',T,D,'g') 

\noindent ylabel('Concentration of $u3, v3, E3$ and $w3$')

\noindent xlabel('Dimensionless Time($t$)') 

\noindent axis square  
 

\begin{thebibliography}{1}
    \bibitem{Briggs1925} G. E. Briggs and J. B. S. Haldane, \emph{A note on the kinetics of enzyme action}. Biochem. J, (1925). 19, 338-339.
\bibitem{Dawkins2007} P. Dawkins, \emph{Differential Equations}.

\noindent http://tutorial.math.lamar.edu/terms.aspx (2007).
  \bibitem{Flach2010} E. H. Flach, S. Schnell, \emph {Stability of open pathways}. Mathematical Biosciences 228 (2010) 147-152.
  \bibitem{Goeke2012}  A. Goeke, Ch. Schilli, S. Walcher and  E. Zerz, \emph {Computing quasi-steady state reductions}. J Math Chem (2012) 50:1495–1513.
   \bibitem{Gorban2011} A.N. Gorban and M. Shahzad,\emph{The Michaelis-Menten-Stueckelberg Theorem}. Entropy (2011), 13, 966-1019. (arXiv:1008.3296 [physics.chem-ph]).
   \bibitem{Gorban2010} A. N. Gorban, O. Radulescu, A. Y. Zinovyev,\emph{Asymptotology of chemical reaction networks}. Chemical Engineering Science 65 (2010) 2310-2324 (arXiv:0903.5072 [physics.chem-ph]).
\bibitem{Hanson2008} S. M. Hanson, S. Schnell, \emph{ Reactant stationary approximation in enzyme kinetics}. J. Phys. Chem. A (2008), 112, 8654-8658.
\bibitem{He1999}  Ji. H. He, \emph {Homotopy perturbation technique}. Computer methods in applied mechanics and engineering 178 (1999) 257-262.
\bibitem{Kargi2009} F. Kargi, \emph { Generalized rate equation for single-substrate enzyme catalyzed reactions}. biochemical and biophysical research communications 382 (2009) 157-159.
 \bibitem{Li2008} B. Li, Y. Shen and B. Li, \emph{Quasi–Steady–state Laws in Enzyme Kinetics}. J. Phys. Chem. A. (2008), 112,  2311-2321.
 \bibitem{Maheswari2011} M. Uma Maheswari, L. Rajendran, \emph{ Analytical solution of non-linear enzyme reaction equations arising in mathematical chemistry}. J Math Chem (2011) 49:1713–1726.
 \bibitem{Meena2010} A. Meena, A. Eswari, L. Rajendran, \emph {Mathematical modeling of enzyme kinetics reaction mechanisms and analytical solutions of non-linear reaction equations}. J Math Chem (2010) 48:179–186.
\bibitem{Murray1989} J. D. Murray, \emph {Mathematical Biology}. Springer, Berlin, (1989), p. 109 .
 \bibitem{Pedersen2008}  M. G. Pedersen, A.  M. Bersani, E. Bersani and G. Cortese, \emph {The total quasi-steady-state approximation for complex enzyme reactions}.Mathematics and Computers in Simulation 79 (2008) 1010–1019.
 \bibitem{Rubinow1975}  S. I. Rubinow, \emph {Introduction to Mathematical Biology}. Wiley, Newyork (1975).
\bibitem{Schnell2000} S. Schnell, P. K. Maini,\emph{ Enzyme Kinetics at High Enzyme Concentration}. Bulletin of Mathematical Biology (2000) 62, 483-499.
\bibitem{Segel1980} L. A. Segel, \emph {Mathematical Models in Molecular and Cellular Biology}. Cambridge University Press, Cambridge(1980).
\bibitem{Varadharajan2011} G. Varadharajan, L. Rajendran, \emph{ Analytical Solutions of System of Non-Linear Differential Equations in the Single-Enzyme, Single-Substrate Reaction with Non-Mechanism-Based Enzyme Inactivation}. Applied Mathematics, (2011) 2, 1140-1147.
\bibitem{Varadharajan2011D} G. Varadharajan, L. Rajendran, \emph {Analytical solution of coupled non-linear second order reaction differential equations in enzyme kinetics}. Natural Science Vol.3, No.6, 459-465 (2011)
\bibitem{Varadharajan2011DD} G. Varadharajan, L. Rajendran,\emph {Analytical solution of coupled non-linear second order reaction differential equations in enzyme kinetics}. Natural science, Vol.3, No.6, 459-465 (2011).

\end{thebibliography}
 \end{document}